  \documentclass[10pt, conference, letterpaper]{IEEEtran}
    \IEEEoverridecommandlockouts
    \makeatletter
    \def\ps@headings{%
    \def\@oddhead{\mbox{}\scriptsize\rightmark \hfil \thepage}%
    \def\@evenhead{\scriptsize\thepage \hfil \leftmark\mbox{}}%
    \def\@oddfoot{}%
    \def\@evenfoot{}}
    \makeatother
\ifCLASSINFOpdf
  \usepackage[pdftex]{graphicx}
  \graphicspath{{../pdf/}{../jpeg/}}
  \DeclareGraphicsExtensions{.pdf,.jpeg,.png}
\else
 \usepackage[dvips]{graphicx}
  \graphicspath{{../eps/}}
  \DeclareGraphicsExtensions{.eps}
\fi

\usepackage{subfigure}

\usepackage[caption=false]{subfig}

\usepackage{amsmath}
\usepackage{amsfonts}

\begin{document}

%
% paper title
% can use linebreaks \\ within to get better formatting as desired
\title{Bifocal-Lens Antenna Based OAM \\ Communications System}
%多模态毫米涡旋波通过双焦距透镜汇聚
\author{\IEEEauthorblockN{Shanghua Gao, Wenchi Cheng, \emph{Senior Member}, \emph{IEEE}, Wei Zhang, \emph{Fellow}, \emph{IEEE}, and Hailin Zhang, \emph{Member}, \emph{IEEE}}~\\[0.2cm]
\vspace{-10pt}

\IEEEauthorblockA{ % State Key Laboratory of Integrated
%Services Networks, Xidian University, Xi'an, China\\
E-mail: \{\emph{shgao@stu.xidian.edu.cn}, \emph{wccheng@xidian.edu.cn}, \emph{w.zhang@unsw.edu.au}, \emph{hlzhang@xidian.edu.cn}\}}

\vspace{-20pt}

\thanks{
Part of this work has been accepted by the IEEE/CIC International Conference on Communications in China (ICCC), Qingdao, China, 2017~\cite{ICCC2017_OAM_con}.

Shanghua Gao, Wenchi Cheng, and Hailin Zhang are with the State Key
Laboratory of Integrated Services Networks, Xidian University, Xi'an, 710071,
China.

Wei Zhang is with the University of New South Wales, Sydney, Australia.
}
}

% use for special paper notices
%\IEEEspecialpapernotice{(Invited Paper)}

% make the title area
\maketitle

\begin{abstract}
Orbital angular momentum (OAM) based radio vortex wireless communications have received much attention recently because it can significantly increase the spectrum efficiency. The uniform circular array (UCA) is a simple antenna structure for high spectrum efficiency radio vortex wireless communications. However, the OAM based electromagnetic waves are vortically hollow and divergent, which may result in the signal loss. Moreover, the divergence of corresponding OAM based electromagnetic wave increases as the order of OAM-mode and radius of UCA increases. Therefore, it is difficult to use high-order OAM-mode, because the corresponding received signal-to-noise ratio (SNR) is very small. To overcome the difficulty of high-order OAM modes transmission, in this paper we propose a lens antenna based electromagnetic waves converging scheme, which maintains the angular identification of multiple OAM-modes for radio vortex wireless communications. We further develop a bifocal lens antenna to not only converge the electromagnetic wave, but also compensate the SNR loss on traditional electromagnetic waves. Simulation results show that the proposed bifocal lens can converge the OAM waves into cylinder-like beams, providing an efficient way to increase the spectrum efficiency of wireless communications.

\end{abstract}

\vspace{10pt}


%\IEEEpeerreviewmaketitle

\section{Introduction}
\IEEEPARstart{D}{ue} to the rapid development of wireless communications, it is now very crowded within the available radio spectrum. Traditionally, the wireless communication was built on the linear momentum based plane-electromagnetic (PE) wave~\cite{6824752,7306545}. However, PE wave has not only the linear momentum, but also the orbital angular momentum (OAM), which is the result of a signal possessing helical phase fronts. The OAM-based radio vortex waves can support multiple orthogonal states/modes, which can be utilized to significantly increase the spectrum efficiency (SE) of wireless communications~\cite{cheng2018orbital,ICCC2017_OAM_hop,GC17_OAM,Thid2007Utilization,8225634}.

There exist a number of antenna structures to generate OAM-based radio vortex beams. A design of flat-lensed spiral phase plate (SPP) was proposed to generate OAM beams~\cite{Bennis2013Flat,Niemiec2014Characterization}. The impedance-matched reflectivity SPP improved the reflectivity of SPP by more than 20 $\rm dB$~\cite{Hui2015Ultralow}.
The authors of~\cite{Wang2015Simple} proposed a parabolic reflecting antenna with an azimuthally deformed Cassegrain subreflector to generate arbitrary OAM-modes. However, these schemes cannot simultaneously generate multiple OAM beams with different OAM-modes. In order to increase SE of wireless communications, multiple OAM-modes need to be generated and transmitted within the same frequency band at the same time. The uniform circular array (UCA), through changing the phase difference of feeds~\cite{Deng2013Generation}, is an effective antenna structure to generate multiple OAM-modes within the same frequency band at the same time~\cite{6711931,7968418,7880700,Wei2015Generation}.
% Applying orbital angular momentum (OAM) beams in low-frequency radio is an efficient way to improve the

Recently, it was shown that the electromagnetic (EM) wave with UCA generated OAM are vortically hollow and divergent~\cite{Wang2015Simple}. Moreover, the EM wave becomes more and more divergent as the order of OAM-mode increases~\cite{5345758}, which leads to very low received signal-to-noise ratio (SNR) for high-order OAM waves~\cite{Mahmouli20134}. Although, the divergent angle decreases as the radius of UCA increases, it is not practical to greatly increase the radius of UCA.
%We discovered the divergence angle decreases with the radius increases. However, This degree of decrease of divergence angle can not satisfy the needs of wireless communication.
In order to facilitate the radio vortex wireless communications, it is desirable to control the OAM beams into a small radiation range. The SPP can generate converged single mode OAM beam~\cite{Chen2016A} while the multi-layer amplitude-phase-modulated surfaces can transform a quasi-spherical wave into non-diffractive single mode OAM beam~\cite{1882-0786-10-1-016701}. However, to the best of our knowledge, no effective method has been proposed to generate multiple simultaneous converged OAM beams. In this work, we focus on the study of generating multiple OAM beams with different OAM-modes while maintaining a small divergent angle all OAM-modes.

\begin{figure*}
\centering
\vspace{5pt}
\includegraphics[width=0.8\textwidth]{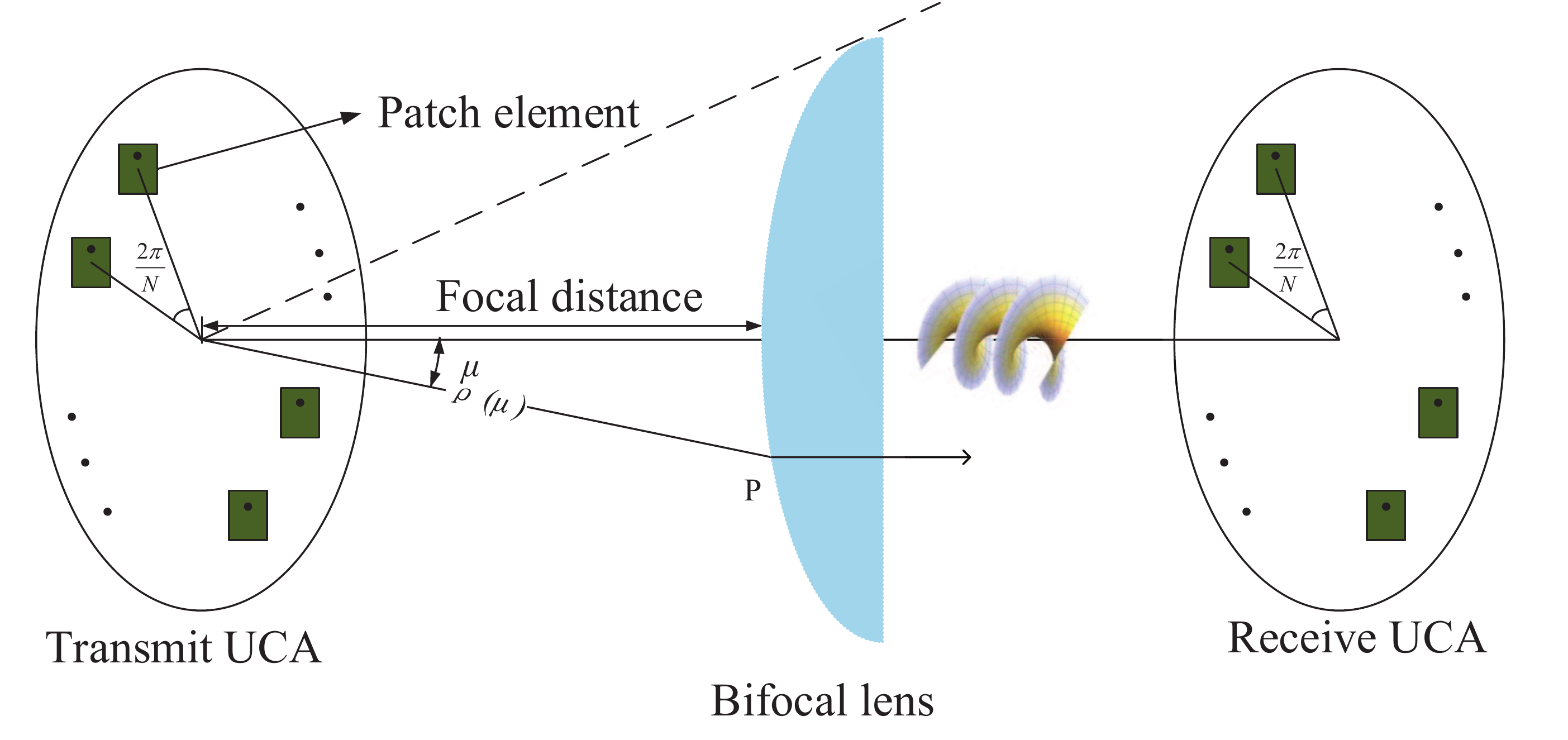}
\caption{The lens-assisted radio vortex wireless communications model.} \label{fig:Lens_System_Model}
\vspace{-10pt}
\end{figure*}

In fact, the lens has the ability of convergence~\cite{1984mgh..book.....J}, which can be used for converging OAM beams. In addition, the lens has no inherent frequency bandwidth limitation, which can be easily extended to many kinds of application scenarios. In this paper, we propose to use the UCA and lens-based antenna to converge multiple modes OAM beams. To avoid the severe loss of the strength of PE beams and OAM beams with low-order OAM-mode due to the large thickness in the center of the lens, we propose a bifocal lens, which has the internal and external parts with different focal distance. In addition, the internal lens has a longer focal distance than the external lens, mitigating the SNR downgrading for PE wave, i.e., OAM wave with mode 0. The converged beams maintain the original wavefront angles. Conventional lens reduces the loss by changing the thickness of the lens. It ensures the refraction direction of PE wave remains the same~\cite{1984mgh..book.....J}. Due to the diversity of OAM divergence angles, we use a novel strategy which uses different focal distances for OAM beams with different OAM-modes.

Our contributions can  be summarized as follows:
\\(1) We reveal that the divergence angle of OAM beams is related to the radius of UCA. Besides, we give the relationships between the radius of UCA and the divergence angle of OAM beams.
\\(2) We show that the lens doesn't change the wavefront angle of OAM beams, which is very crucial for orthogonal multiplexing in wireless communications.
\\(3) We propose the bifocal lens to not only converge OAM beams but also compensate the SNR loss on PE waves and low-order OAM beams.
 %However, plane beam faces a bigger attenuation for the center of lens antenna has the biggest thickness. For this, our proposed bifocal lens antenna is divided into internal lens and external lens with different focal distance. The two parts of bifocal lens antenna are distinguished by different divergence angles, thus gives high OAM modes much greater convergence effect and reduce the attenuation of plane beam. Internal lens has a longer focal distance compared to external lens, thus it can be thinner than the single focal lens, which provides plane beam with a smaller decay. In addition, the side lobes of the beam are aliased with other modes. To solve this problem, for high mode OAM beams, the external lens does not cover the divergence angle of the side lobes. For plane beam, the lens antenna has the ability of reducing the power of the feed to the edge, which reduce the the influence of side lobes.
 %因为毫米波相比光波有更大的波动性，所以波束较宽.所以如果一个不同模态的涡旋波被单焦距镜汇聚

The rest of this paper is organized as follows. Section~\ref{sec:sys} describes the system model of lens-assisted converged radio vortex wireless communications. Section~\ref{sec:III}, and~\ref{sec:IIII}, and~\ref{sec:V} design and analyze UCA antenna, lens antenna, and bifocal lens antenna, respectively. We also show the lens antenna does not change the wavefront of OAM beams. Section~\ref{sec:simu} evaluates our developed lens converging and the spectrum efficiency enhancement.
%描述产生并汇聚多模态毫米波段涡旋波的系统模型。    公式化并解出符合涡旋波多模态不同汇聚状态的制镜者公式
The paper concludes with Section~\ref{sec:conc}.

\section{The System Model}\label{sec:sys}
\begin{figure}[!]
\centering
%\vspace{3pt}
\includegraphics[scale=0.40]{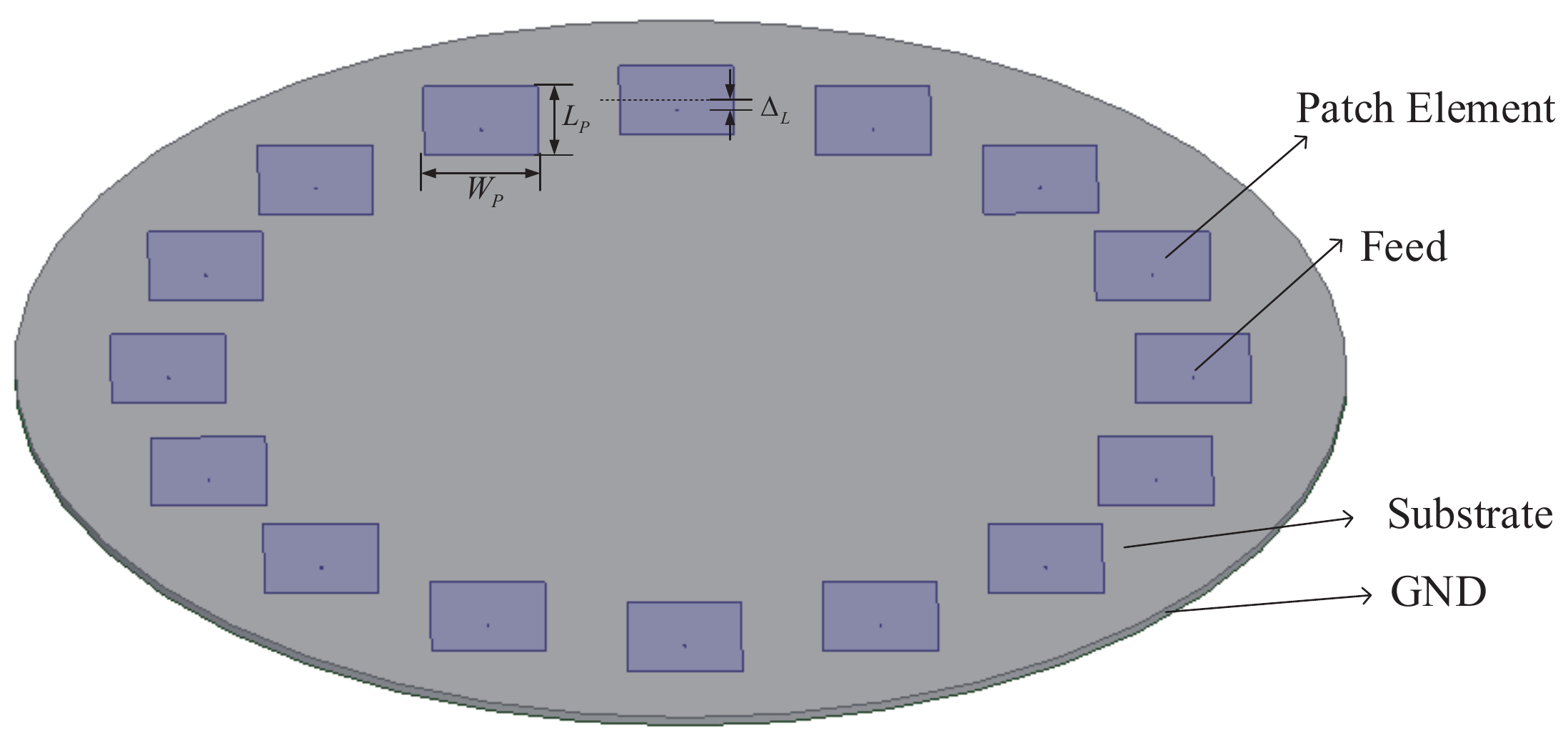}
\caption{The profile of 16 patch elements UCA with 16 patch elements.} \label{fig:UCA}
\vspace{-5pt}
\end{figure}
We consider the lens-assisted radio vortex wireless communications model, as shown in Fig.~\ref{fig:Lens_System_Model}, where the transmitter and receiver are both UCA antennas with $N$ array-elements equally distributed on the circle. By controlling the phase information, UCA with $N$ elements is designed to simultaneously generate multiple OAM beams with different OAM-modes. The phase information of adjacent element is linearly increased by $2\pi l/N$, where $l$ is the index/order of OAM-mode. The maximum index of OAM-mode that UCA can support is limited by $-N/2   \leq l < N/2$~\cite{Niemiec2014Characterization}. The PE beam (OAM wave with mode 0) is generated when all elements of UCA have the same phase information. The UCA generated OAM waves are vortically hollow and divergent, depending on the radius of UCA and the index of OAM-mode~\cite{5345758}. The lens antenna is designed based on the principle of wave path consistency~\cite{1984mgh..book.....J}. The OAM beams can be treated as emitting from the central point of the transmit UCA. We place the UCA at the focal point of lens antenna. Multiple divergent OAM beams generated by UCA are sent to lens antenna, where P is the incident point the beam entered the lens, and $\mu$ is the angle between the beam transmitting direction and the central axis. Lens antenna converges these OAM beams into cylinder-like beams while maintaining the original wavefront of each OAM-mode. Therefore, the receive antenna can efficiently receive the OAM beams within a small reception area.

%As the diameter of UCA increases, the OAM beam becomes more and more convergent while the side lobes increase. It's tricky to balance those two factors.5.In order to generate converged multiple OAM beams with small side lobes, we use the lens antenna to converge the divergent OAM beams generated by UCA.

\section{Analsysis of Unconverged OAM Beams} \label{sec:III}
In this section, we design and analyze the UCA antenna. We analyze the divergence angle of OAM beams. We reveal that the divergence angle of OAM beams is related to the radius of UCA. Also, we give the relationships between the radius of UCA and the divergence angle of OAM beams. We then derive the capacity for OAM based radio vortex wireless communications with unconverged OAM beams.
\subsection{Design of UCA Antenna}
\par We design the patch based UCA antenna. The profile of a 16 patch elements based UCA is shown in Fig.~\ref{fig:UCA}, where patch elements are equally distributed around the circle. To achieve the best feed performance, it is needed to design the size of patch elements~\cite{Milligan1985Modern}.
For the specific design of UCA antenna, please refer to the APPENDIX of this paper.
\begin{table*}
\caption{The simulated relationship between the radius of UCA and the divergent angle of OAM beams.}
\centering
\begin{tabular*}{16cm}{cccccccccccccccc}
\hline
$R$ (mm) &8.8 & 9.9 & 11.0 & 12.1 & 13.2 & 14.3 & 15.4 & 16.5 & 17.6 & 18.7 & 19.8 & 20.9 & 22.0 & 23.1 & 24.2 \\
\hline
$\theta_1$ (degree) & 16.4 & 14.7 & 12.9 & 12.1 & 10.9 & 9.6 & 8.7 & 8.3 & 8.1 & 8 & 7.6 & 7.1 & 6.6 & 6.1 & 5.8\\
$\theta_2$ (degree) & 27.7 & 25.3 & 21.5 & 19.2 & 17.5 & 16.6 & 15.4 & 14 & 12.8 & 12.2 & 12 & 11.8 & 11.3 & 10.5 & 9.9\\
$\theta_3$ (degree) & 38.4 & 34.7 & 29 & 26.9 & 25.1 & 22 & 20.1 & 19.2 & 18.9 & 18.3 & 17.2 & 16.2 & 15 & 14.2 & 13.5\\
$\theta_4$ (degree) & 57 & 44.3 & 41 & 33.8 & 30.5 & 29.3 & 28.3 & 24.5 & 22.5 & 21.8 & 21.4 & 21.1 & 19.9 & 18.5 & 17.3\\
\hline
\label{tab:R_theta}
\end{tabular*}
\end{table*}

%\subsection{Divergence angle of OAM Based Vortex Beams}
\subsection{Characteristics of OAM Based Vortex Beams}
\par
According to~\cite{Johan2009Angular}, the E field of OAM beams can be expressed as follows:
\begin{eqnarray}
 E(r,\theta,\varphi)=A(r)e^{il\varphi}J_{l}(2kR\sin\theta),
\label{eq:ezE_of_OAM}
\end{eqnarray}
where $r$ represents the distance between beams and axis, $\varphi$ is azimuth, $l$ is the index of OAM-mode, $J_l(\cdot)$ is Bessel function, $\theta$ is the angle between the normal to the plane of the circle and the direction from UCA to the field, and $A(r)$ is the amplitude of OAM beams. $R$ is the radius of UCA, which is the distance from the center of transmit UCA to the center of patch elements. The phase rotation $\exp(i l \varphi)$ controls the spatial distribution of OAM beam. The expression for $A(r)$ can be given as follows~\cite{Johan2009Angular}:
\begin{eqnarray}
A(r)=-j\frac{\mu_0\omega d}{4\pi}Ni^{-l}\frac{e^{ikr}}{r},
\label{eq:A_r}
\end{eqnarray}
where $j$ represents the current density of the electric dipole, $\omega$ is wavelength, $k=\omega / c$, $\mu_0$ is the magnetic conductivity, and $d$ is the length of electric dipole.

\par We denote by $\theta_{l}$ the divergent angle of the $l$th OAM-mode, where the corresponding OAM beam has the largest gain. We notice that the number of patch elements causes little impact on the divergent angle $\theta_{l}$.
%Thus, we ignore the influence of $N$ on $\theta_{l}$.
By simulating using HFSS~\cite{ansys2015ansoft}, we get the relationship between $R$ and $\theta$, as shown in Table~\ref{tab:R_theta}. $\theta_{l}$ decreases as $R$ increases, while the descent rate of $\theta_{l}$ decreases as $R$ increases. Also, $\theta_{l}$ increases as OAM-mode increases. Analyzing the obtained data in Table~\ref{tab:R_theta}, we propose the following two mathematical models to characterize the relationship between $R$ and $\theta_l$.
\par Model 1: The first model is given as follows:
\begin{eqnarray}
\theta_l= a_{l} R^{b_{l}}
\label{eq:theta_R_power},
\end{eqnarray}
where $a_{l}$ is the factor that controls the magnitude of $\theta_{l}$, while $b_{l}$ is the factor that controls the descent rate of curves. To accommodate the descent rate, $b_{l}$ should be a negative number.
\par Model 2: The second model is given as follows:
\begin{eqnarray}
\theta_l=\frac{p_{l}}{(R+q_{l})},
\label{eq:theta_R_rational}
\end{eqnarray}
where $p_{l}$ mainly controls the the magnitude of $\theta_{l}$ and $q_{l}$ controls is a negative number controlling the descent rate of curves.
%\begin{figure}[htbp]
%\centering
%\includegraphics[scale=0.33]{simulation_theta.pdf}
%\caption{The simulated relationship between the radius of UCA-$R$ and the divergence angle of OAM beams-$\theta_l$.} %\label{fig:R_theta}
%\end{figure}
\par
\begin{figure}[htbp]
		\subfigure[$\theta_l= a_{l} R^{b_{l}}$]{
			 \includegraphics[width=9cm,height=7cm]{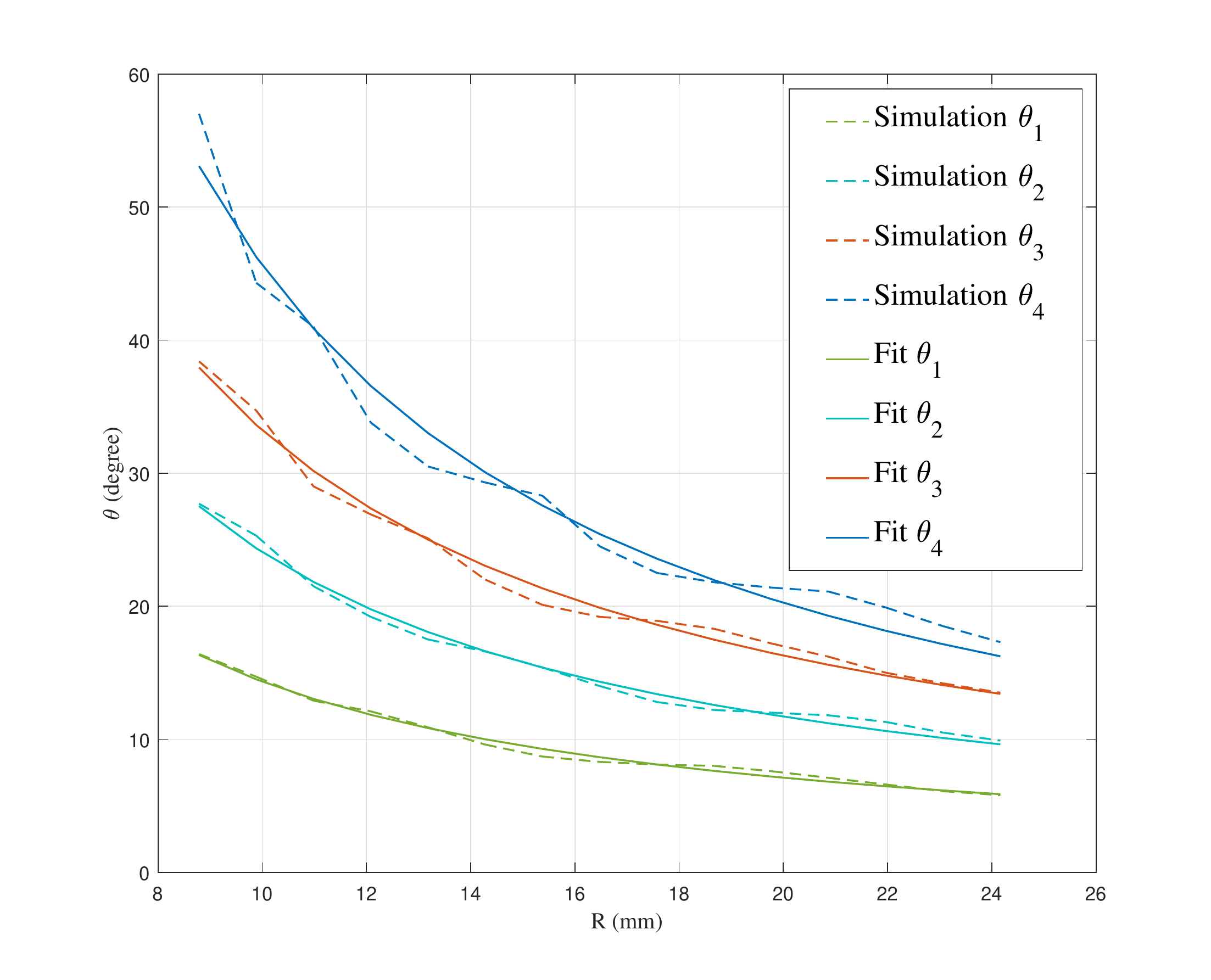}
			}
		 \subfigure[$\theta_l=\frac{p_{l}}{(R+q_{l})}$]{
			 \includegraphics[width=9cm,height=7cm]{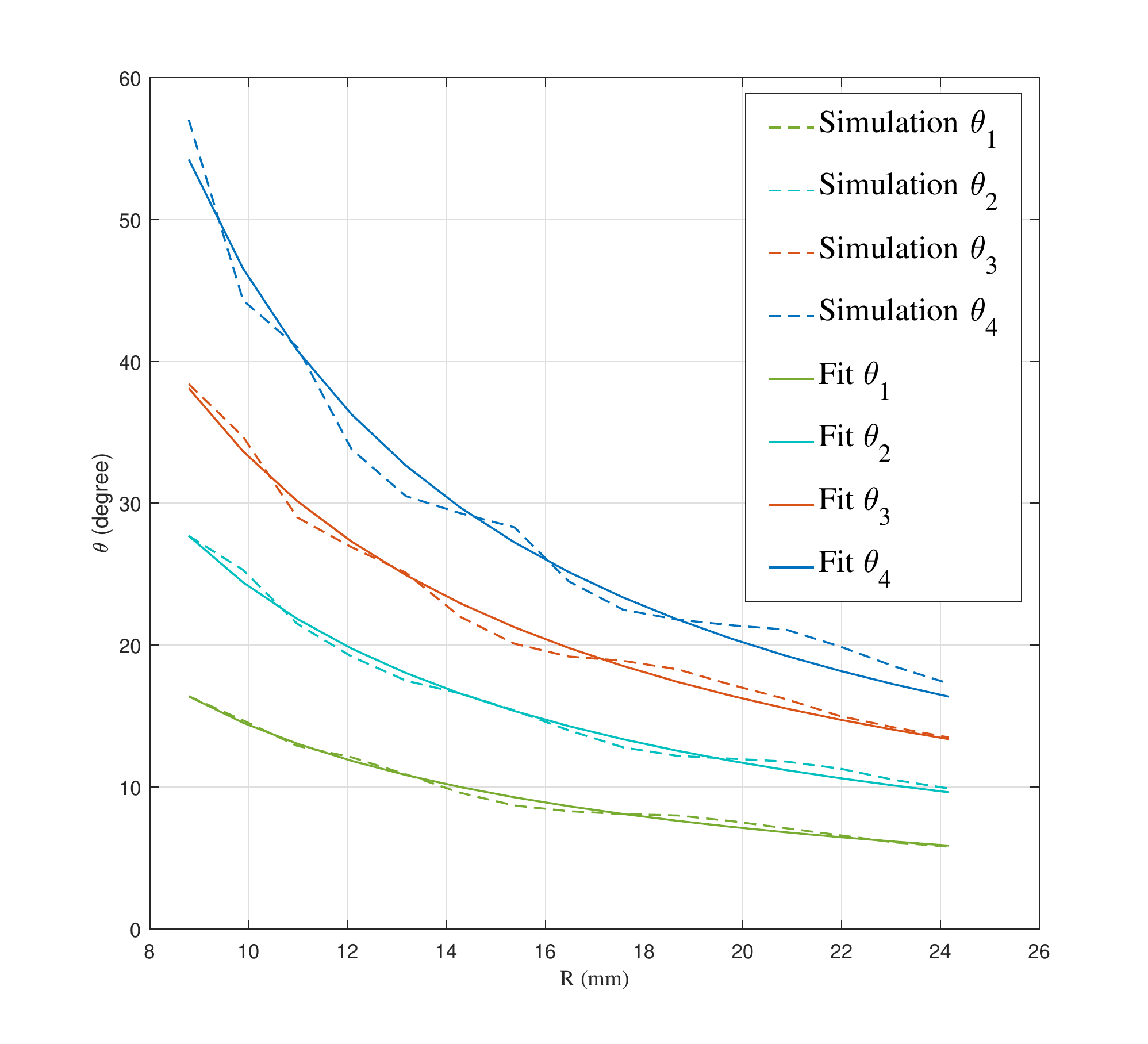}
			}
   \caption{The relationship between the radius of UCA and the divergent angle of OAM beams.} \label{fig:R_theta_fit}
\end{figure}

 Figure~\ref{fig:R_theta_fit} gives the relationship between the radius of UCA and the divergent angle of OAM beams. The curves in Fig.~\ref{fig:R_theta_fit}(a) indicate the relationship between $\theta_l$ and $R$ is in the form of $\theta_l= a_{l} R^{b_{l}}$. The curves in Fig.~\ref{fig:R_theta_fit}(b) are in the form of $\theta_l=\frac{p_{l}}{(R+q_{l})}$. The specific values of $a_{l}$,$b_{l}$, $p_{l}$, and $q_{l}$ that obtained by simulation is shown in Table.~\ref{theta_R_spec}. As shown in Fig. \ref{fig:R_theta_fit}, our proposed two curves can fit the relationship between $\theta_l$ and $R$ quite well. The divergent angle increases as the increase of OAM-mode. In addition, the slope of divergent angle decreases as the radius of UCA increases.
\begin{table}
\caption{Specific values of $a_{l}$,$b_{l}$, $p_{l}$ and $q_{l}$.}
\begin{tabular*}{8cm}{ccccc}
\hline
OAM-mode & $a_{l}$ & $b_{l}$ & $p_{l}$ & $q_{l}$ \\
\hline
1& 147   & -1.011 & 140.9 &-0.1902\\
2& 263.2 & -1.039 & 227.2 &-0.5844\\
3& 354.3 & -1.028 & 317.1 &-0.4647\\
4& 676.3 & -1.171 & 360.7 &-2.135 \\
\hline
\label{theta_R_spec}
\end{tabular*}
\end{table}

\subsection{Capacity analysis of OAM Based Radio Vortex Wireless Communications}
The received power of receive antenna, denoted by $P_r$, can be given as follows:~\cite{Kraus2003Antennas}:
\begin{eqnarray}
P_r=\frac{G_{t} A_{er}}{4 \pi d^{2}}P_t,
\label{eq:pr_pt}
\end{eqnarray}
where $P_t$ is the power of transmit antenna, $G_{t}$ is the gain of transmit antenna, $d$ represents the distance between the receive antenna and transmit antenna, and $A_{er}$ represents the effective reception area when the antenna is pointed to the maximum direction of the beam.
%We focus on the design of the transmit antenna in this paper. Thus,
We assume that the radius of receive antenna is constant, denoted by $r_0$, and the transmit antenna is aligned with the receive antenna. The effective reception area of the receive antenna can be written as follows:
%, for the plane beam, we assume that the receive antenna is always aligned to the maximum direction of the incoming beam. For the OAM beams,
\begin{eqnarray}
A_{er}=\frac{\lambda^2 G_0}{4 \pi},
\label{eq:Aer}
\end{eqnarray}
where $G_0$ is the gain of receive antenna.
%For the plane beam, the receiving SNR can be written as:

%{\rm SNR}=\frac{P_r}{N}=\frac{{G_{t}}^{0} A_{er}}{(4 \pi d)^{2}N}P_t=\frac{{G_{t}}^{0} \lambda^2 G_0}{(4 \pi d)^{2}N}P_t,

%where $N$ is the noise mean power, ${G_{t}}^{0}$ is the gain of transmit antenna.
For OAM beams, each diverged OAM beam has a divergent angle with the value of $2\theta_{l} \pm 2 \Delta\theta_{l}$, where $\Delta\theta_{l}$ is the half-power beamwidth of the OAM beam corresponding to the $l$th mode.
Since the effective reception area depends on the propagation distances, we discuss the following three cases.
\setcounter{equation}{9}
\begin{figure*}[!htb]
\begin{eqnarray}
C=\sum_{l=1}^L B \log_{2}{(1+{\rm SNR}_{l})}=
  \left\{
   \begin{array}{lll}
      \sum_{l=1}^L B \log_{2}{\left(1+{\frac{{G_{t}}(l) \lambda^2 G_0}{(4 \pi d)^{2}N_{0}}P_t}\right)}
       \quad d< \frac{r_0}{\tan(\theta_{l} +  \Delta\theta_{l})};\\
      \sum_{l=1}^L B \log_{2}{\left(1+{\frac{{G_{t}}(l,\theta) \lambda^2 G_0}{(4 \pi d)^{2}N_{0}}P_t}\right)}
      \quad \frac{r_0}{\tan(\theta_{l} +  \Delta\theta_{l})}<d<\frac{r_0}{\tan(\theta_{l} -  \Delta\theta_{l})};\\
      0 \quad  d> \frac{r_0}{\tan(\theta_{l} -  \Delta\theta_{l})}.
   \end{array}
   \right.
\label{eq:capacity_before}
\end{eqnarray}
\hrulefill
\end{figure*}
\setcounter{equation}{6}
\par Case 1: The receive antenna covers the outside of the half-power beam $2\theta_{l} + 2 \Delta\theta_{l}$ when the power of main lobe can be fully received. The corresponding SNR and the region of $d$ are given as follows:
\begin{eqnarray}
  \left\{
   \begin{array}{lll}
   {\rm SNR}_{l}=\frac{P_r}{N_{0}}=\frac{{G_{t}}(l) A_{er}}{(4 \pi d)^{2}N_{0}}P_t=\frac{{G_{t}}(l) \lambda^2 G_0}{(4 \pi d)^{2}N_{0}}P_t;\\
  d< \frac{r_0}{\tan(\theta_{l} +  \Delta\theta_{l})},\\
   \end{array}
   \right.
\label{eq:SNR1}
\end{eqnarray}
where ${G_{t}}(l)$ is the gain when the transmit antenna emits the $l$th OAM-mode.
%\begin{eqnarray}
%d=\frac{r_0}{\tan(\theta +  \Delta\theta_{l})}.
%\label{eq:d1}
%\end{eqnarray}

\par Case 2: The receive antenna cannot cover the inside of the half-power beam $2\theta_{l} - 2 \Delta\theta_{l}$ when the power of main lobe cannot be received. In this case, we have $P_r\rm=\rm0$ and thus we have
\begin{eqnarray}
  \left\{
   \begin{array}{lll}
     {\rm SNR}_{l}=0;\\
     d> \frac{r_0}{\tan(\theta_{l} -  \Delta\theta_{l})}.\\
   \end{array}
   \right.
\label{eq:SNR2}
\end{eqnarray}

\par Case 3: The receive antenna partially receives the beams.
We denote by ${G_{t}}(l,\theta)$ the maximum power of the $l$th mode OAM beams that can be received from the direction of $\theta$. Under this scenario, the SNR and the region of $d$ can be obtained as follows:
\begin{eqnarray}
  \left\{
   \begin{array}{lll}
      {\rm SNR}_{l}=\frac{P_r}{N_{0}}=\frac{{G_{t}}(l,\theta) A_{er}}{(4 \pi d)^{2}N_{0}}P_t=\frac{{G_{t}}(l,\theta) \lambda^2 G_0}{(4 \pi d)^{2}N_{0}}P_t;\\
     \frac{r_0}{\tan(\theta_{l} +  \Delta\theta_{l})}<d<\frac{r_0}{\tan(\theta_{l} -  \Delta\theta_{l})}.\\
   \end{array}
   \right.
\label{eq:SNR3}
\end{eqnarray}

%The distance range between the receive antenna and the transmit antenna is:
%\begin{eqnarray}
%\frac{r_0}{\tan(\theta +  \Delta\theta_{l})}<d<\frac{r_0}{\tan(\theta -  \Delta\theta_{l})}.
%\label{eq:d3}
%\end{eqnarray}
Now, we can derive the capacity of the OAM based radio vortex wireless communications using divergent OAM beams indexed from 1 to $L$~\cite{WCpp}, denoted by $C$, in Eq. \eqref{eq:capacity_before}.
${\rm SNR}_{l}$ is the SNR of radio vortex wireless communication using the $l$th divergent OAM-mode and $B$ is system bandwidth.
In case 1, when the propagation distance is short, the main lobe of unconverged beams can be fully received. In case 2, when the propagation distance is relatively long, the main lobe of beams can only be partly received. Thus, the capacity decreases faster. In case 3, when the propagation distance is very large, the beams cannot be received by the antenna. Thus, the obtained capacity is very close to 0. Since multiple OAM-modes can be utilized together, the capacity increases as more and more OAM-modes (especially the high-order OAM-modes) are used. Because $F_r$ and $\lambda$ are in reciprocal relationship. The operating frequency is also related to the effective reception area of receiving antenna, the capacity increases as operating frequency increases.

\section{Analysis of Lens Antenna} \label{sec:IIII}
In this section, we design and analyze the lens antenna.  We analyze the phase and amplitude of OAM beams converged by the lens. We show that the lens doesn't change the wavefront angle of OAM beams, which is very crucial for orthogonal multiplexing in wireless communications. We then derive the capacity of lens-converged OAM based radio vortex wireless communications.
\subsection{Design of Lens Antenna}

%When generating OAM beams by UCA，
 %\begin{eqnarray}
% \varphi=\frac{2\pi}{N}
%\label{eq:phi_uca}
%\end{eqnarray}
%where $N$ represents the number of elements in UCA. In our module, $N$=16.
%where $A$$(r)$ is the amplitude of beams,$r$ 表示到波束中心轴线的辐射距离,$\varphi$ 为方位角,$l$ 是轨道角动量的本征值\\
%相位旋转因子$exp(il\varphi)$决定了涡旋波束空间相位分布结构，不同ＯＡＭ模态涡旋波束的空间结构不同\\
%利用UCA生成涡旋波时，
 %\begin{eqnarray}
 %\varphi=\frac{2\pi}{N}
%\label{eq:phi_uca}
%\end{eqnarray}
%其中$N$代表UCA振子数，在本文中$N$=16
%轨道角动量应用在电磁波中，在正常电磁波中添加一个相位旋转因子$exp(il\varphi)$,
%使电磁波波前结构不在是平面结构，而是绕着波束传播方向旋转，呈现出一种螺旋的相位结构，
%涡旋电磁波可表示为：\\
%$U(r,\varphi)=A(r)e^{il\varphi}$
\par According to Fermat's principle, the wave path from the UCA through the lens to the aperture plane needs to be equalized~\cite{Milligan1985Modern}. The wave path along the axis and the wave path at any angle denoted as $\mu$ are equal when reaching the aperture plane of the lens, as shown in  Fig.~\ref{fig:Lens_System_Model}. The lensmaker's equation in polar coordinates is given as follows~\cite{Silver1984Microwave,Johnson1961Antenna}:
%运用费马定理，使从馈源通过透镜到达口径面的光程相等。如图.~\ref{fig:Lens_2d}所以令轴上焦点到透镜平面的距离与任意角度发射波到透镜平面的距离相等。~\cite{Milligan1985Modern}易得制镜者公式:
\setcounter{equation}{10}
\begin{eqnarray}
 \rho(\mu)=\frac{(n-1)f}{n\cos\mu-1},
\label{eq:lenmaker_origin}
\end{eqnarray}
%其中,f代表焦距，n代表折射率，$\mu$代表距离主轴的角度
where $f$ represents the focal distance, $n = \sqrt{\varepsilon_r\mu_r}$ is the index of refraction with $\varepsilon_r$ and $\mu_r$ representing the relative permittivity and permeability of the lens medium, respectively.

Substituting $\rho(\mu)^2=(x+f)^2+y^2$ and $\rho(\mu)=f+nx$ into Eq. \eqref{eq:lenmaker_origin}, the lensmakers equation in cartesian coordinates can be derived as follows:
 \begin{eqnarray}
 (n^2-1)x^2+2(n-1)fx-y^2=0,
\label{eq:lenmaker_single}
\end{eqnarray}
where (x, y) is the coordinates of point $P$.
It can be seen that the lensmaker's equation is hyperbolic with a progressive line. Thus, the maximum divergent angle that the lens antenna can support can be obtained as follows:
%可以看出制镜者公式是一条双曲线，双曲线有渐进线，以此可以求得透镜最大支持的发散角度
 \begin{eqnarray}
\mu_{\rm max}=\arccos\frac{1}{n}.
\label{eq:mu_max}
\end{eqnarray}
This angle can be used to verify whether the designed lens antenna covers the divergent angle of OAM beams or not.
We denote by $\theta_{\rm max}$ the maximum divergent angle of OAM beams. Thus, we have $\theta_{\rm max}<\mu_{\rm max}$. Then,
%此角度在仿真时可以验证设计透镜是否能覆盖涡旋波的发散角度.
the diameter, denoted by $D$, for the lens antenna can be derived as follows:
 \begin{eqnarray}
 D=2\rho(\mu) \sin\theta_{\rm max}=\frac{2(n-1)f\sin\theta_{\rm max}}{n\cos\theta_{\rm max}-1}.
\label{eq:D_max}
\end{eqnarray}
Moreover, in order to limit the lens size while maintaining a low downgrading, we need to design the values of $D$ and $f$. We denote by $m$ the coefficient that balances $D$ and $f$. Then, we have
\begin{eqnarray}
 D=mf.
\label{eq:D_balance}
\end{eqnarray}
Combining Eq.~\eqref{eq:D_max} and Eq.~\eqref{eq:D_balance}, we can obtain $m$ as follows:
\begin{eqnarray}
 m=\frac{2(n-1)\sin\theta_{\rm max}}{n\cos\theta_{\rm max}-1}.
\label{eq:m}
\end{eqnarray}

For Model 1 and Model 2 corresponding to Eqs. \eqref{eq:theta_R_power} and \eqref{eq:theta_R_rational}, respectively, $m$ can be re-written as:
\begin{eqnarray}
 m=\frac{2(n-1)\sin(a_{j} R^{b_{j}})}{n\cos(a_{j} R^{b_{j}})-1},
\label{eq:R_m_power}
\end{eqnarray}
and
\begin{eqnarray}
 m=\frac{2(n-1)\sin \left[  \frac{p_{j}}{(R+q_{j})} \right] }{n\cos \left[\frac{p_{j}}{(R+q_{j})} \right]-1},
\label{eq:R_m_rational}
\end{eqnarray}
respectively.

%So far, $\theta_{\rm max}$ can only be obtained by simulation.
%\subsection{Characteristics of Converged OAM Beams}
\par Next, we show that converging does not change the wavefront of OAM beams. We give the relationships between the phase change of beam and its passing wave path as follows~\cite{Clapp1965OPTICAL}:
 \begin{eqnarray}
\Delta\varphi(r)=k L(r),
\label{eq:phrase_light_length}
\end{eqnarray}
where $r^2=x^2+y^2$ and
%其中k为光波的波速，$k=\frac{2\pi}{\lambda}$ 式中$\lambda$为波长
$L(r)$ is the wave path.

Observing Eq.~\eqref{eq:phrase_light_length}, we find that the beam passing through the lens causes phase changes and the lens has a phase modulation for the incident wave.
Because the design of the lens antenna uses the wave path consistent principle, the wave path $L(r)$ between transmit UCA and lens antenna's aperture plane is constant regardless of the variation of $r$. We denote by $\Delta L_p$ the wave path.
Ignoring the impact of the lens on the amplitude of incident wave and combining Eq. \eqref{eq:ezE_of_OAM}, the E field of OAM beams after passing through the lens antenna can be derived as follows:
\begin{eqnarray}
E^{'}(r,\theta,\varphi)&=&A(r)J_{l}(2kR\sin\theta)e^{il\varphi+i\Delta\varphi(r)}\nonumber \\
&=&A(r)J_{l}(2kR\sin\theta)e^{il\varphi+ik\Delta L_p} \nonumber \\
&=&e^{ik\Delta L_p}A(r)J_{l}(2kR\sin\theta)e^{il\varphi}.
\label{eq:U_after_lens_final}
\end{eqnarray}
%Because the design of the lens antenna uses the optical path consistent principle, the wave path $L(r)$ between source and lens antenna's aperture plane is constant regardless of the variation of $r$. We denote by $\Delta L$ the wave path. Thus, Eq. \eqref{eq:U_after_lens} can be rewritten as:
%\begin{eqnarray}
%E^{'}(r,\theta,\varphi)&=&A(r)J_{l}(2kR\sin\theta)e^{il\varphi+ik\Delta L} \nonumber \\
%&=&e^{ik\Delta L}A(r)J_{l}(2kR\sin\theta)e^{il\varphi}.
%\label{eq:U_after_lens_final}
%\end{eqnarray}
Therefore, after passing through the lens antenna, the phases of OAM wave corresponding to different angles only increased by a fixed value $\Delta\varphi$. Thus, our designed lens antenna maintains the wavefront characteristics of OAM beams while converging the OAM beams into cylinder-like waves.
\par The lens changes the amplitude distribution of beams on the antenna aperture surface~\cite{Johnson1961Antenna}. The feed power is equal to the caliber power in the ideal scenario. However, due to the divergence of OAM beams, there exits power loss since the size of lens is limited. The amplitude of beams after passing through the lens, denoted by $A_{L}^{'}(r,f)$, is given as follows:
\begin{eqnarray}
A_{L}^{'}(r,f)=A(r)\frac{a(n\cos \varphi -1)^{3}}{f^2(n-1)^{2}(n-\cos \varphi)},
\label{eq:E_after_lens_1}
\end{eqnarray}
where $a$ is the ratio of the energy entering the lens to the total energy.
In addition, the downgrading caused by the lens is proportional to the thickness of the lens. The amplitude of beams after passing through the lens, denoted by $A_{L}(r,f)$ can be written as~\cite{769344}:
\begin{eqnarray}
A_{L}(r,f)=A_{L}^{'}(r,f)-pT(f,\theta),
\label{eq:E_after_lens}
\end{eqnarray}
where $p$ is the downgrading factor related to material of lens given range $1 \sim 10$ \cite{1984mgh..book.....J} and $T(f)$ is the thinness of lens given as follows:
\begin{eqnarray}
 T(f,\theta)=-\frac{f}{n+1}+\sqrt{\bigg(\frac{f}{n+1}\bigg)^2+\frac{\left(\frac{D}{2}-f \tan{\theta}\right)^2}{n^2-1}}.
\label{eq:T_of_lens}
\end{eqnarray}
\subsection{Capacity analysis of lens-converged OAM Based Radio Vortex Wireless Communications}
\setcounter{equation}{24}
\begin{figure*}[htbp]
\begin{eqnarray}
C_{L}=\sum_{l=1}^L B \log_{2}{(1+{\rm SNR}_{l}^{'})}=\sum_{l=1}^L B \log_{2}{\left\{1+{\frac{\left[G_{t}^{'}(l)\frac{a(n\cos \varphi -1)^{3}}{f_0^2(n-1)^{2}(n-\cos \varphi)}-pT(f,\theta_{l})\right] \lambda^2 G_0}{(4 \pi d)^{2}N_{0}}P_t}\right\}}.
\label{eq:capacity_converge}
\end{eqnarray}
\hrulefill
\end{figure*}

After converging, OAM beams are almost converged into cylinder-like beam. However, the converged OAM beams still have the small divergent angle, denoted by $\sigma$. By substituting Eq. \eqref{eq:E_after_lens_1} into Eq. \eqref{eq:SNR1}, the maximum propagation distance of beams, denoted by $d_{\rm max}$, and SNR of lens based radio vortex wireless communication using converged OAM with the index $l$, denoted by ${\rm SNR}_{l}^{'}$ can be written as follows:
%For the converged plane beam, the SNR can be written as:
%{\rm SNR}=\frac{P_r}{N}=\frac{{G_{t}}^{0'} A_{er}}{(4 \pi d)^{2}N}P_t=\frac{{G_{t}}^{0'} \lambda^2 G_0}{(4 \pi d)^{2}N}P_t,
%where ${G_{t}}^{0'}$ is the gain of bifocal lens antenna.
\setcounter{equation}{23}
\begin{eqnarray}
  \left\{
   \begin{array}{lll}
      {\rm SNR}_{l}^{'}=\frac{\left[G_{t}^{'}(l)\frac{a(n\cos \varphi -1)^{3}}{f^2(n-1)^{2}(n-\cos \varphi)}-pT(f,\theta_{l})\right] A_{er}}{(4 \pi d)^{2}N_{0}}P_t \\
        \qquad  \ \  =\frac{\left[G_{t}^{'}(l)\frac{a(n\cos \varphi -1)^{3}}{f^2(n-1)^{2}(n-\cos \varphi)}-pT(f,\theta_{l})\right] \lambda^2 G_0}{(4 \pi d)^{2}N_{0}}P_t;\\
     d_{\rm max}=\frac{r_0}{\tan \sigma},\\
   \end{array}
   \right.
\label{eq:SNR3_2}
\hrulefill
\end{eqnarray}
where $G_{t}^{'}(l)$ is the gain when the transmit antenna (including the UCA antenna and the lens antenna) emits the converged $l$th mode OAM beams.

The capacity of converged OAM beams can be written in Eq. \eqref{eq:capacity_converge}, where $C_{L}$ is capacity of radio vortex wireless communication using converged OAM with the index from 0 to $L$.

\section{Analysis of Bifocal Lens Antenna} \label{sec:V}
In this section, we design and analyze the bifocal lens antenna. We propose the bifocal lens to not only converge OAM beams, but also compensate the SNR loss on PE waves and low-order OAM beams. We then derive the capacity of bifocal lens OAM based radio vortex wireless communications.
\subsection{Design of Bifocal Lens Antenna}
\begin{figure}
\centering
\vspace{5pt}
\includegraphics[scale=0.48]{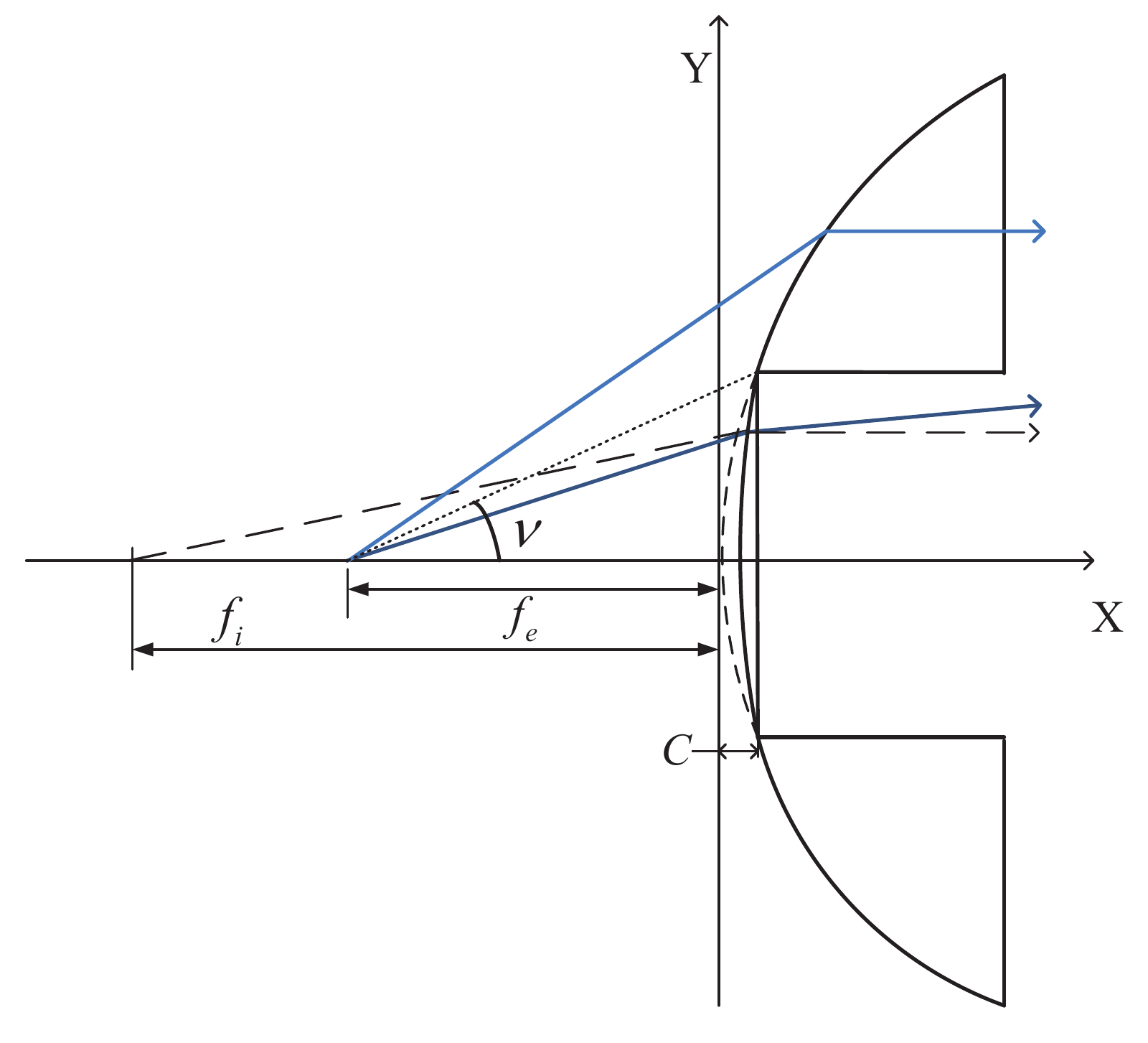}
\caption{Schematic diagram of bifocal lens antenna.} \label{fig:bifocal_lens_theory}
\vspace{-10pt}
\end{figure}

Since the center of lens antenna is thicker than other parts, the strength of PE beam (i.e., OAM beam with mode 0) and OAM beams with low-order OAM-mode severely downgrades after passing through the lens according to Eq. \eqref{eq:E_after_lens}. To solve this problem, we design the bifocal lens antenna, as shown in Fig.~\ref{fig:bifocal_lens_theory}, which is divided into the internal and external parts with different focal distances. The internal lens has a long focal distance while the external lens has a short focal distance. Thus, the bifocal lens can be thinner than the single focal lens in the center area, significantly mitigating the SNR downgrading of PE beam and OAM beams with low-order OAM-mode. %3.Moreover, low mode OAM beams also face the attenuation of lens, which influences the OAM beams transmission. Our method can also be applied to OAM beams while maintain the characteristics of low index of OAM beams.

We denote by $\nu$ the angle corresponding to the boundary of separating the internal lens and the external lens. Since each OAM-mode has an inherent divergent angle of $\theta_{l}$, where the amplitude corresponding to the $l$th OAM-mode has the largest value. To avoid two parts of lens causing different convergence effects on the same OAM-mode, $\nu$ should be selected in the middle of different divergent angles. The internal lens and the external lens share the same spindle. Based on the lensmaker's equation of single focal lens antenna in Eq.~\eqref{eq:lenmaker_single}, we can write the lensmaker's equation for bifocal lens antenna as follows:
%构造双焦距透镜公式，二者共享同一光轴，根据辐射角度不同区分不同的焦距，如Fig.~\ref{fig:single_refraction}所示。现写出双焦距透镜折射公式：\\
\setcounter{equation}{25}
\begin{eqnarray}
  \left\{
   \begin{array}{lll}
   (n^2-1)(z-C)^2+2(n-1)f_i\times(z-C)-x^2=0;\\
   (n^2-1)z^2+2(n-1)f_e \times  z-x^2=0;\\
      \frac{x}{z+f_e}=\tan{\nu};\\
      \theta_l<\nu<\theta_{l+1},
   \end{array}
   \right.
\label{eq:bifocal_lens}
\end{eqnarray}
where $f_{i}$ and $f_{e}$ represent the focal distances of the internal lens and external lens, respectively. Solving Eq.~\eqref{eq:bifocal_lens}, we can obtain the boundary's coordinates of the internal lens and the external lens. In our design, $f_{e}$ is a prefixed value. We need to obtain the value of $f_{i}$ to solve Eq.~\eqref{eq:bifocal_lens}.  The parameter that controls the thickness of the internal lens and impacts the convergence, denoted by $\rho$, is written as follows:
\begin{eqnarray}
f_{i}=\rho f_{e},
\label{eq:f2}
\end{eqnarray}
where $\rho > 1$.
%为内部透镜和外部透镜的分界角度，由计算和仿真测试得到最优分界值.n为折射率
 %Since each mode of the OAM beams have an inherent divergence angle of ${\mu_a}^{n}$, where the intensity of beam is greatest. In the range of the internal lens, when OAM beam pass through the lens, if the direction of angle ${\mu_a}^{n}$ meets the principle of optical path consistency, it can be considered that the convergence of the internal lens still maintains the original OAM-mode characteristics. When optical path between the beam emitted from ${\mu_a}^{n}$ and the beam emitted from the main axis is $m\lambda$, the beams from two directions are consistent at the wavefront of the lens antenna's aperture plane. The focal distance of the internal lens can be obtained:
\par
\begin{figure}
\centering
\vspace{5pt}
\includegraphics[scale=0.5]{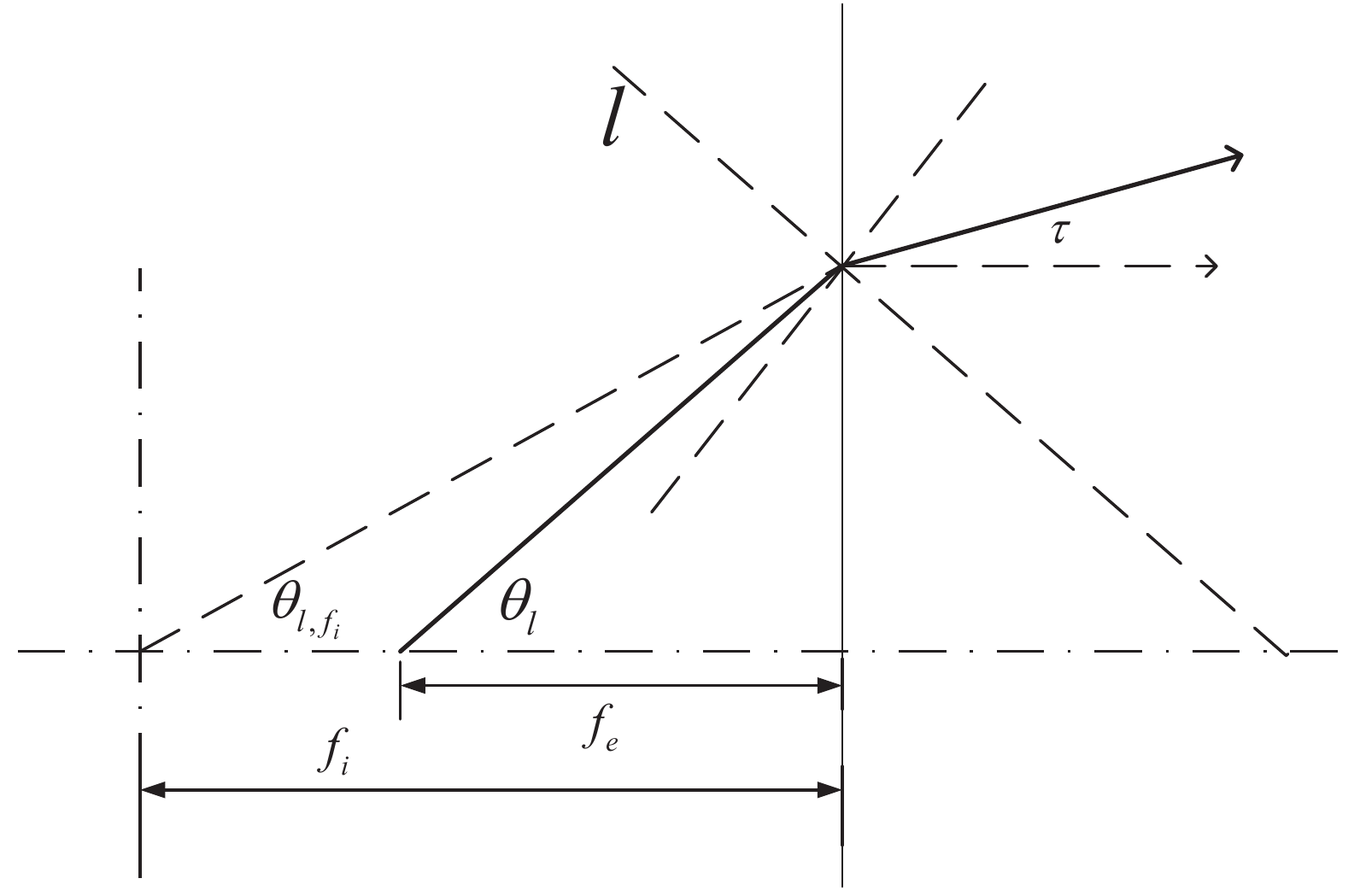}
\caption{Refraction path of internal lens.} \label{fig:bifocal_inside}
\vspace{-10pt}
\end{figure}

  To obtain $\rho$, we need to make sure that the internal lens has the similar convergence effect compared with the external lens. Fig.~\ref{fig:bifocal_inside} shows the refraction path of internal lens, where the solid line indicates the path of OAM beams that emitted from focal $f_e$. In the range of the internal lens, due to the relatively long focal distance of the internal lens, OAM beams still have a small divergent angle, denoted by $\tau$, after passing through the lens, which can be easily obtained by Fig.~\ref{fig:bifocal_inside}:
\begin{eqnarray}
\tau=\arcsin \left[\frac{1}{n}\sin(\theta_l+\theta_t)\right]-\theta_t,
\label{eq:mu_a_ans}
\end{eqnarray}
where
\begin{eqnarray}
\theta_t=\arctan \left\{ \frac{\sin(\theta_{l,f_i})}{n-\cos(\theta_{l,f_i})} \right \} .
\label{eq:theta_l}
\end{eqnarray}
However, if the direction of original divergent angle meets the principle of wave path consistency, the $l$th mode OAM beam still maintains the characteristics of the $l$th mode OAM beams after passing through the lens.
Specifically, if wave path is equal to $m\lambda$, where $m$ is an positive integer, the phase of beams remains unchanged. Therefore, we need to make sure that the difference of wave path between the internal lens and the single focal lens equals to $m\lambda$.

\par
According to Fig.~\ref{fig:bifocal_inside}, the wave path emitted from the focal $f_e$, denote by $L(f_e)$, is given as follows:
\begin{eqnarray}
L(f_e)=\frac{f_e}{\cos\theta_{l}}+\frac{nT_{\rm max}}{\cos\tau},
\label{eq:fe_orign}
\end{eqnarray}
where $T_{\rm max}$ is maximum thickness of the internal lens, and $T_{\rm max}/{\cos\tau}$ is the thickness that beams pass through the internal lens. On contrast, as the focal distance of lens is $f_i$, the beam that emitted from the focal $f_i$ has no divergent angle after passing through the lens. Therefore, the wave path emitted from focal $f_i$, denote by $L(f_i)$, is given as follows:
\begin{eqnarray}
L(f_i)=\frac{f_i}{\cos{(\theta_{l,f_i}})}+nT_{\rm max},
\label{eq:fi_orign}
\end{eqnarray}
where $\theta_{l,f_i}$ is the divergent angle of OAM beams when the focal distance is $f_i$.
\par
If the difference between $L(f_i)$ and $L(f_e)$ is equal to $m\lambda$, the wavefront of beams that emitted from focal $f_e$ remains unchanged after passing through the internal lens.
Because the internal lens is very thin, in order to simplify the calculation, $\frac{nt_{max}}{\cos\tau}$ can be approximate to $nt_{\rm max}$. Thus, the simplified relationship between $f_i$ and $f_e$ is given as follows:
\begin{eqnarray}
m\lambda+\frac{f_e}{\cos\theta_{l}}=\frac{f_i}{\cos{(\theta_{l,f_i})}}.
\label{eq:f1_f0_after}
\end{eqnarray}
Combining $f_i\tan{(\theta_{l,f_i})}=f_e\tan\theta_{l}$ and Eq.~\eqref{eq:f1_f0_after}, $f_i$ that maintaining the wavefront of OAM beams is obtained as follows:
\begin{eqnarray}
f_{i}=\sqrt{\left( m\lambda+\frac{f_e}{\cos\theta_{l}}\right) f_e\tan{\theta_{l}}}.
\label{eq:f0_final}
\end{eqnarray}
The downgrading caused by lens on OAM beams of high-order OAM-mode are relatively small. To simply the design of bifocal lens, we only make the internal lens cover the range of $1$th mode OAM beams.
Substituting Eq.~\eqref{eq:f2} into Eq.~\eqref{eq:f0_final}, $\rho$ can be written as follows:
\begin{eqnarray}
\rho=\frac{\sqrt{\left( m\lambda+\frac{f_e}{\cos\theta_{1}}\right)f_e\tan{\theta_{1}}}}{f_e}.
\label{eq:rho_final}
\end{eqnarray}

%双焦距透镜的外部分对涡旋波的影响在上一部分已推出，先考虑内部透镜的作用。

%鉴于每个模态的涡旋波都有一个固有的发散角度$\mu_{an}$，在此角度下，模态n的电场强度最大。在内部透镜的范围内，当n模态的涡旋波经过透镜时，角度$\mu_{an}$的方向
%保持光程一致性原则，则可看做内部透镜的汇聚依然可以近似保证涡旋波的模态特征。当波束从$\mu_{an}$发射与沿主轴发射的光程相差$m$个波长，则两方向在透镜平面的波前保持一致。
%结合单焦距透镜的仿真结果，本文的内部透镜只汇聚0、1两个模态。0模态等同传统的阵列天线发射的波束。所以只要保证1模态的发散角度$\mu_{a1}$处满足光程一致即可。
In order to better analyze the bifocal lens, we give the amplitude of OAM beams after passing through the bifocal lens. Based on the amplitude of beams after passing through the single focal lens in Eq. \eqref{eq:E_after_lens}, a longer focal distance of the internal lens makes the lens thinner in the center, while a thinner lens reduces the downgrading of OAM beams. When the divergent angle of OAM beams is smaller than $\nu$, the beams pass through the internal lens. Otherwise, the beams pass through the external lens.  Thus, the amplitude of OAM beams after passing through the bifocal lens is given as follows:
\begin{eqnarray}
 A_B(r,\theta)=
  \left\{
   \begin{array}{lll}
     A_{L}(r,f_i)\quad \theta<\nu;\\
     A_{L}(r,f_e) \quad \theta>\nu.\\
   \end{array}
   \right.
\label{eq:E_after_bifocallens}
\end{eqnarray}

%\subsection{Convergence Effect of Internal Lens}

With the amplitude of beams after passing through the bifocal lens of the single focal lens, we can prove that our designs significantly increase the SE of wireless communications. To evaluate our developed lens and bifocal lens, we are going to give the capacity of OAM based radio vortex wireless communications in the next section.
\subsection{The Capacity of bifocal lens OAM Based Radio Vortex Wireless Communications}
The bifocal lens can mitigate the SNR downgrading caused by lens. According to Eq. \eqref{eq:E_after_bifocallens}, SNR of bifocal lens based radio vortex wireless communication using converged OAM with the index $l$, denoted by ${\rm SNR}_{l}^{''}$, can be written as follows:
\begin{eqnarray}
{\rm SNR}_{l}^{''}=
\frac{A_{B}(r,\theta) A_{er}}{(4 \pi d)^{2}N_{0}}P_t.
\label{eq:SNR4}
\end{eqnarray}
When the beam of the lth OAM-mode passes through the internal lens, the internal lens impacts the amplitude of beams, that is $ \theta<\nu$, $A_B(r,\theta)=A_L(r,f_i)$. On the other hand, when the beam of the lth OAM-mode passes through the external lens, the amplitude of beam is impacted by the external lens, that is $ \theta>\nu$, $A_B(r,\theta)=A_L(r,f_e)$. Thus, the capacity of radio vortex wireless communication using OAM beams converged by the bifocal lens, denoted by $C_{B}$, is expressed in Eq. \eqref{eq:capacity_bifcallens}.

\par After converging by the lens antenna, OAM beams are converged into cylinder-like beams. When the propagation distance is short, main lobes of both converged and diverged beams can be received be antenna.
%However, the converged beams are downgraded by lens antenna. The remaining power of beams is still large enough for the wireless communication.converged OAM beams still have a small divergent angle, which is too small to influence the wireless communication. Thus,
When the propagation distance is long enough, the divergent OAM beams cannot be received by antenna. The the converged OAM beams can be received within a relatively long distance. Thus, the converged OAM beams with multiple OAM-modes can be used in a relatively long range to increase the spectrum efficiency of wireless communication.

\begin{figure*}[ht]
\begin{eqnarray}
C_{B}=\sum_{l=1}^L B \log_{2}{(1+{\rm SNR}_{l}^{''})}=
  \left\{
   \begin{array}{lll}
      \sum_{l=1}^L B \log_{2}{\left\{1+{ \frac{\left[G_{t}^{'}(l)\frac{a(n\cos \varphi -1)^{3}}{f_i^2(n-1)^{2}(n-\cos \varphi)}-pT(f_i,\theta_{l})\right] \lambda^2 G_0}{(4 \pi d)^{2}N_{0}}P_t}\right\}}
      \quad \theta<\nu;\\
      \sum_{l=1}^L B \log_{2}{\left\{1+{\frac{\left[G_{t}^{'}(l)\frac{a(n\cos \varphi -1)^{3}}{f_e^2(n-1)^{2}(n-\cos \varphi)}-pT(f_e,\theta_{l})\right] \lambda^2 G_0}{(4 \pi d)^{2}N_{0}}P_t}\right\}}
      \quad \theta>\nu.
   \end{array}
   \right.
\label{eq:capacity_bifcallens}
\end{eqnarray}
\hrulefill
\end{figure*}

\section{Simulation and Evaluation}\label{sec:simu}
In this section, we evaluate the performance of our proposed lens antenna and bifocal lens antenna designs, where HFSS~\cite{ansys2015ansoft} is used to simulate and verify our designs. Moreover, the obtained capacity is verified. We set $f_{r}=35\rm~GHz$ and $\varepsilon_{r} = 2.2$. The parameters of UCA are calculated according to Eqs.~\eqref{eq:patch_W}-\eqref{eq:patch_l} and we have $W_P=3.388\rm~mm$, $L_P=2.947\rm~mm$, $\Delta L=0.438\rm~mm$, and $\varepsilon_{re}=2.039$.
%The parameters of UCA can be found in the APPENDIX of this paper.
%The operating frequency of antenna is set to $35\rm~GHz$. In this paper,

%For UCA, the number of patch elements is associated with the divergent angle of OAM beams. Fig.~\ref{8_16_comparison} shows the divergent angles of beam generated by 8 patch elements UCA and 16 patch elements UCA with the 1st OAM mode. As shown in Fig.~\ref{8_16_comparison}, the divergent angle of 8 patch elements UCA is roughly twice of the divergent angle of 16 patch elements UCA, which means that the large number of patch elements results in the small divergent angle.

According to the relationship between $\theta$ and $R$ mentioned before, we choose to use 16 patch elements UCA in the following simulation by trading off the size of UCA and the divergent angle of OAM beams. We select $f_e = 30 \rm~mm$, $m=1.67$, $R=0.6 \lambda$ and $\rho= 2.17$. Thus, we hace $f_i = 65.3 \rm~mm$ and $r = 25 \rm~mm$, which can be obtained using Eq. \eqref{eq:f2}.

\begin{figure}[htbp]
\centering
\begin{minipage}[t]{0.32\linewidth}
\centering
\centerline{\includegraphics[width=2.9cm,height=5cm]{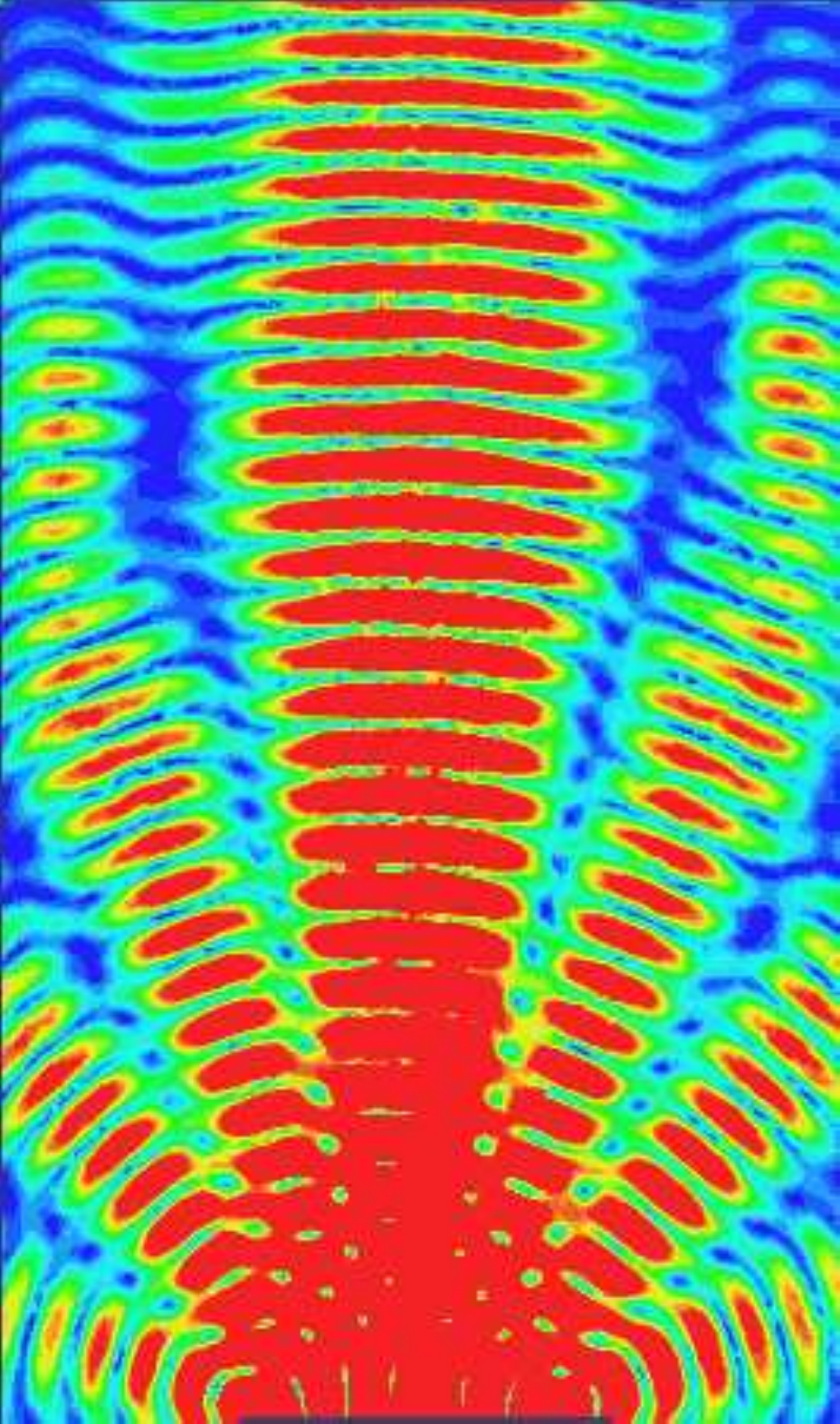}}
\centerline{$l=0$ (PE beam)}
\end{minipage}
\centering
\begin{minipage}[t]{0.32\linewidth}
\centering
\centerline{\includegraphics[width=2.9cm,height=5cm]{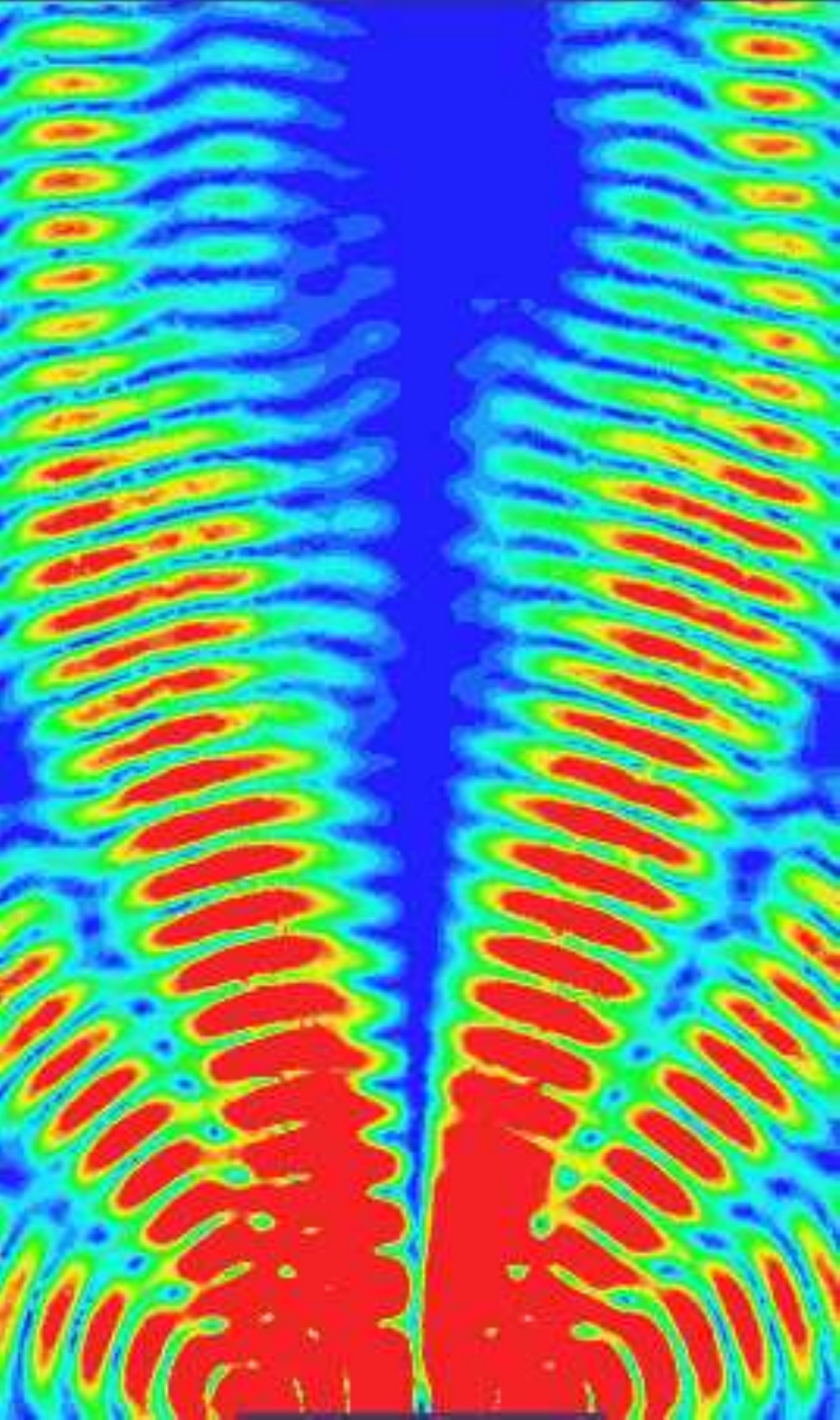}}
\centerline{$l=2$}
\end{minipage}
\centering
\begin{minipage}[t]{0.32\linewidth}
\centering
\centerline{\includegraphics[width=2.9cm,height=5cm]{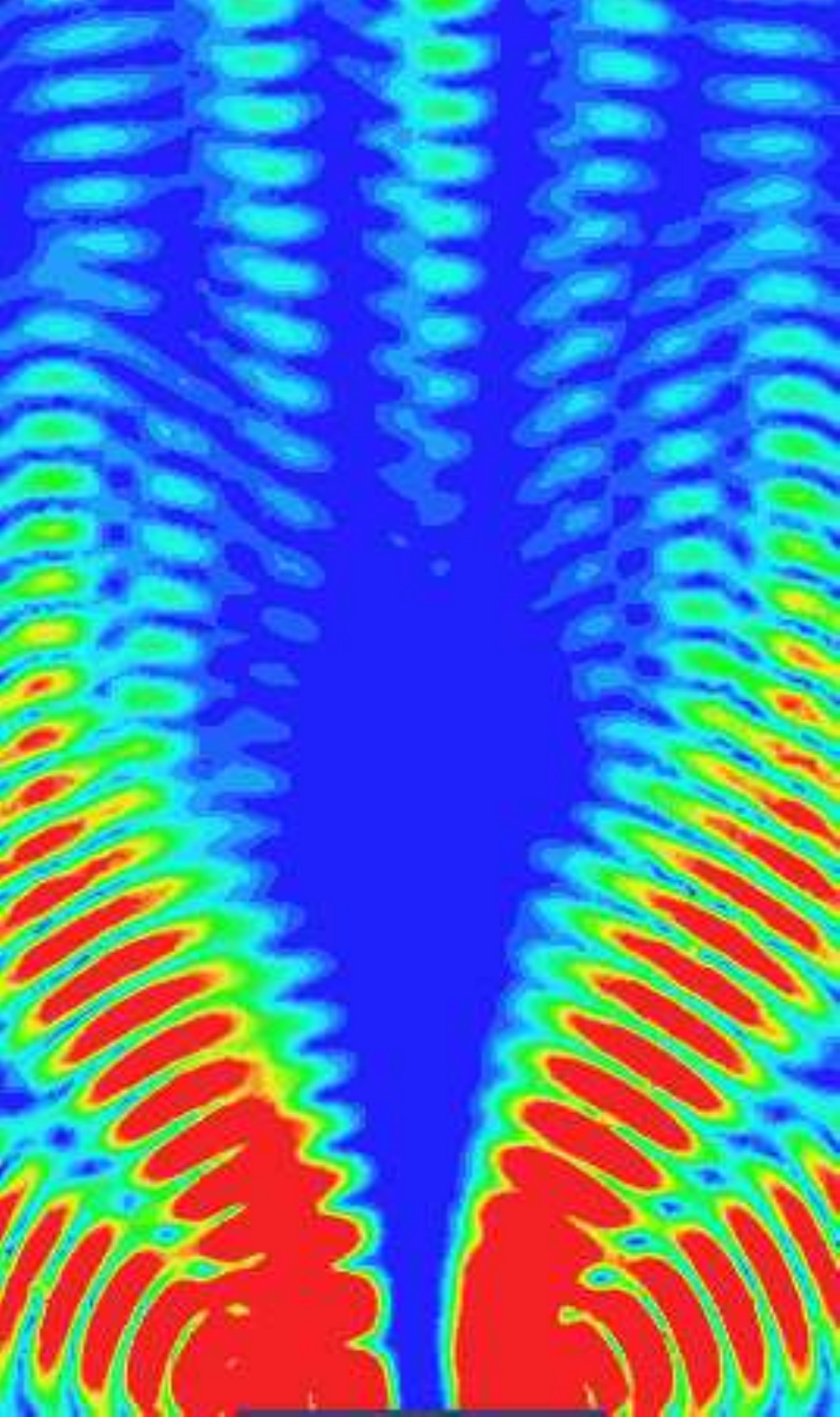}}
\centerline{$l=4$}
\end{minipage}

\centering
\begin{minipage}[t]{0.32\linewidth}
\centering
\centerline{\includegraphics[width=2.9cm,height=5cm]{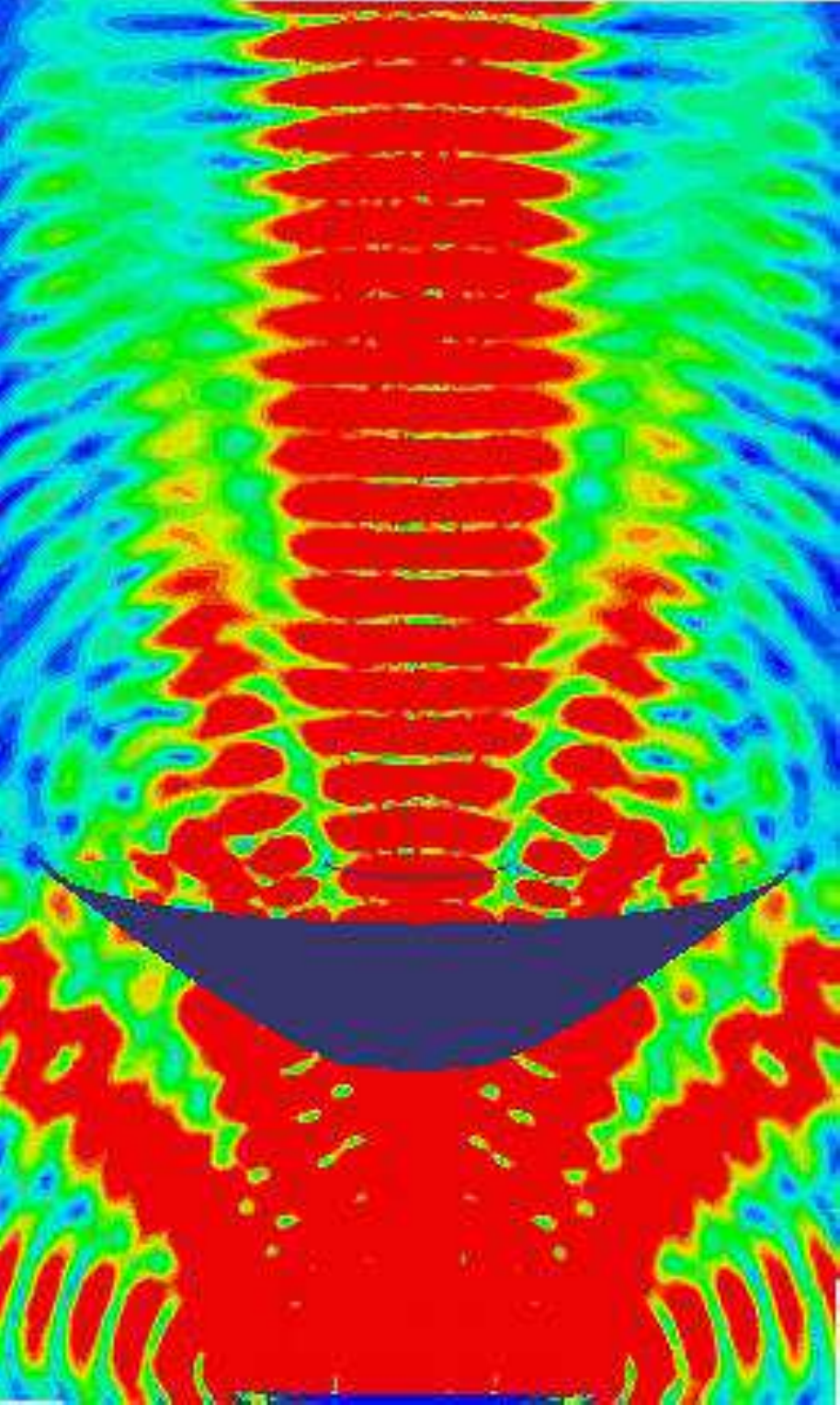}}
\centerline{$l=0$ (PE beam)}
\end{minipage}
\centering
\begin{minipage}[t]{0.32\linewidth}
\centering
\centerline{\includegraphics[width=2.9cm,height=5cm]{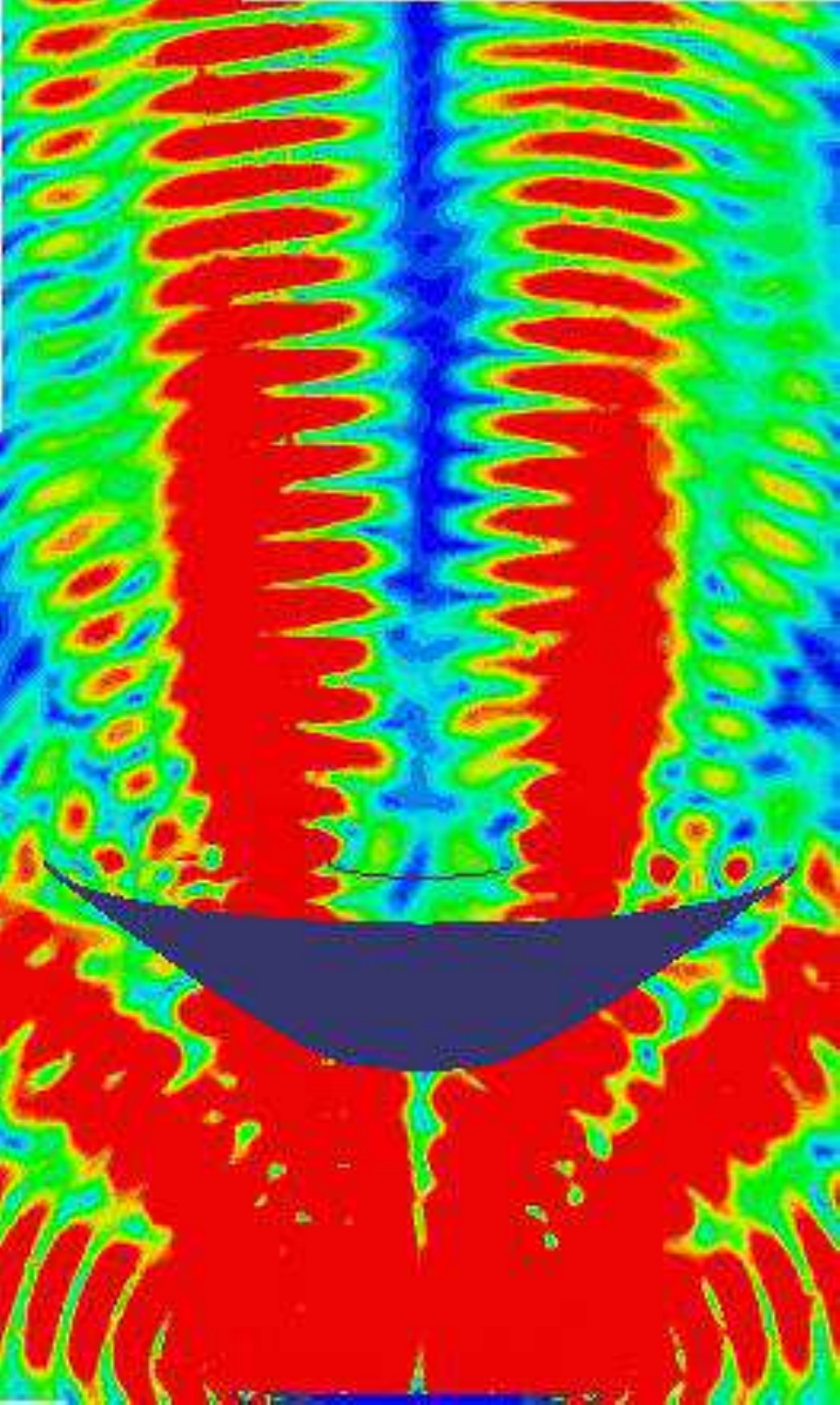}}
\centerline{$l=2$}
\end{minipage}
\centering
\begin{minipage}[t]{0.32\linewidth}
\centering
\centerline{\includegraphics[width=2.9cm,height=5cm]{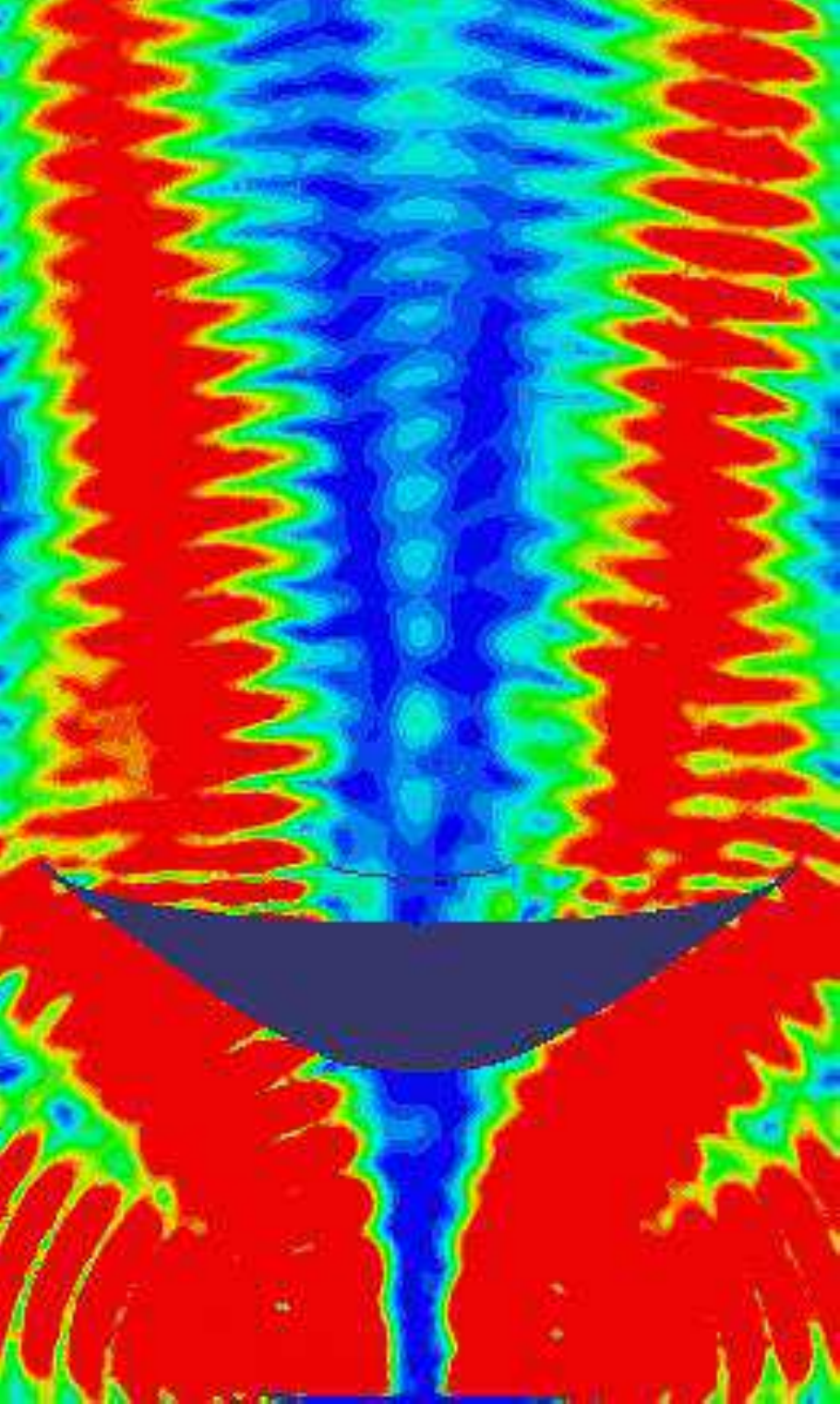}}
\centerline{$l=4$}
\end{minipage}
 \caption{E field of OAM beams (horizontal direction).}\label{E_field}
   \end{figure}

\begin{figure}[htbp]
%\vspace{-13pt}
\centering
\begin{minipage}[t]{0.32\linewidth}
\centering
\centerline{\includegraphics[width=2.8cm,height=2.8cm]{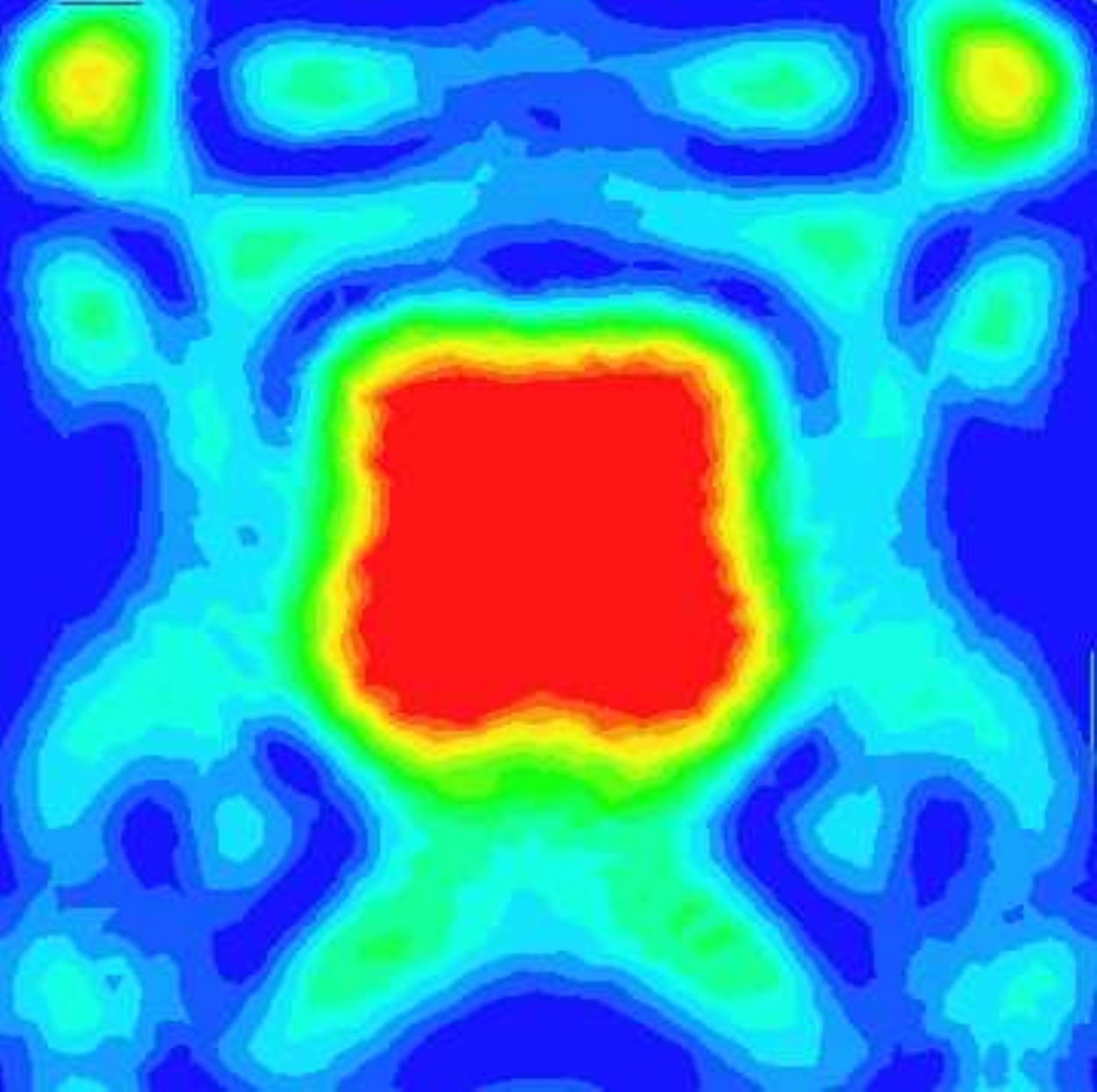}}
\centerline{$l=0$ (PE beam)}
\end{minipage}
\centering
\begin{minipage}[t]{0.32\linewidth}
\centering
\centerline{\includegraphics[width=2.8cm,height=2.8cm]{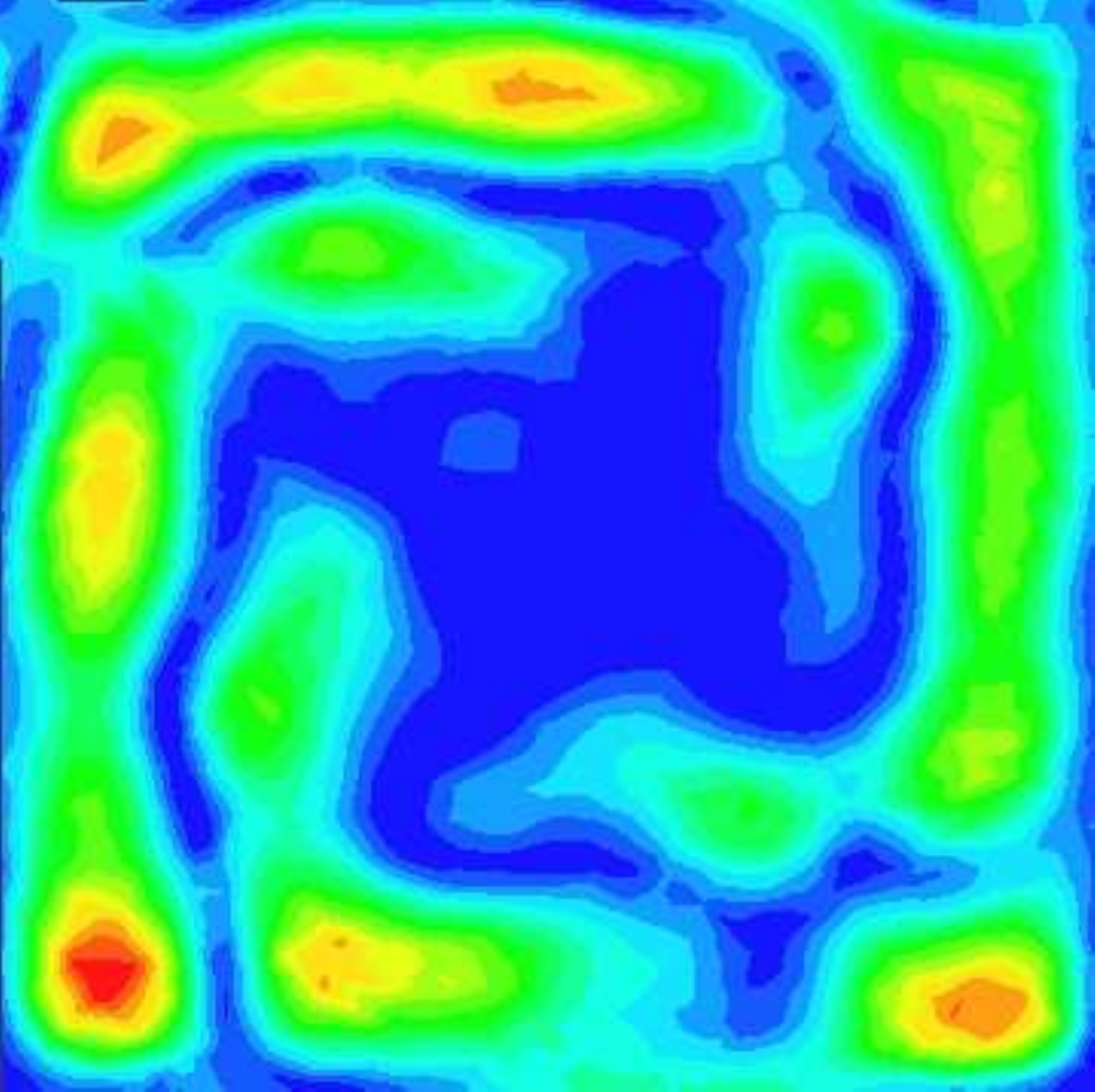}}
\centerline{$l=2$}
\end{minipage}
\centering
\begin{minipage}[t]{0.32\linewidth}
\centering
\centerline{\includegraphics[width=2.8cm,height=2.8cm]{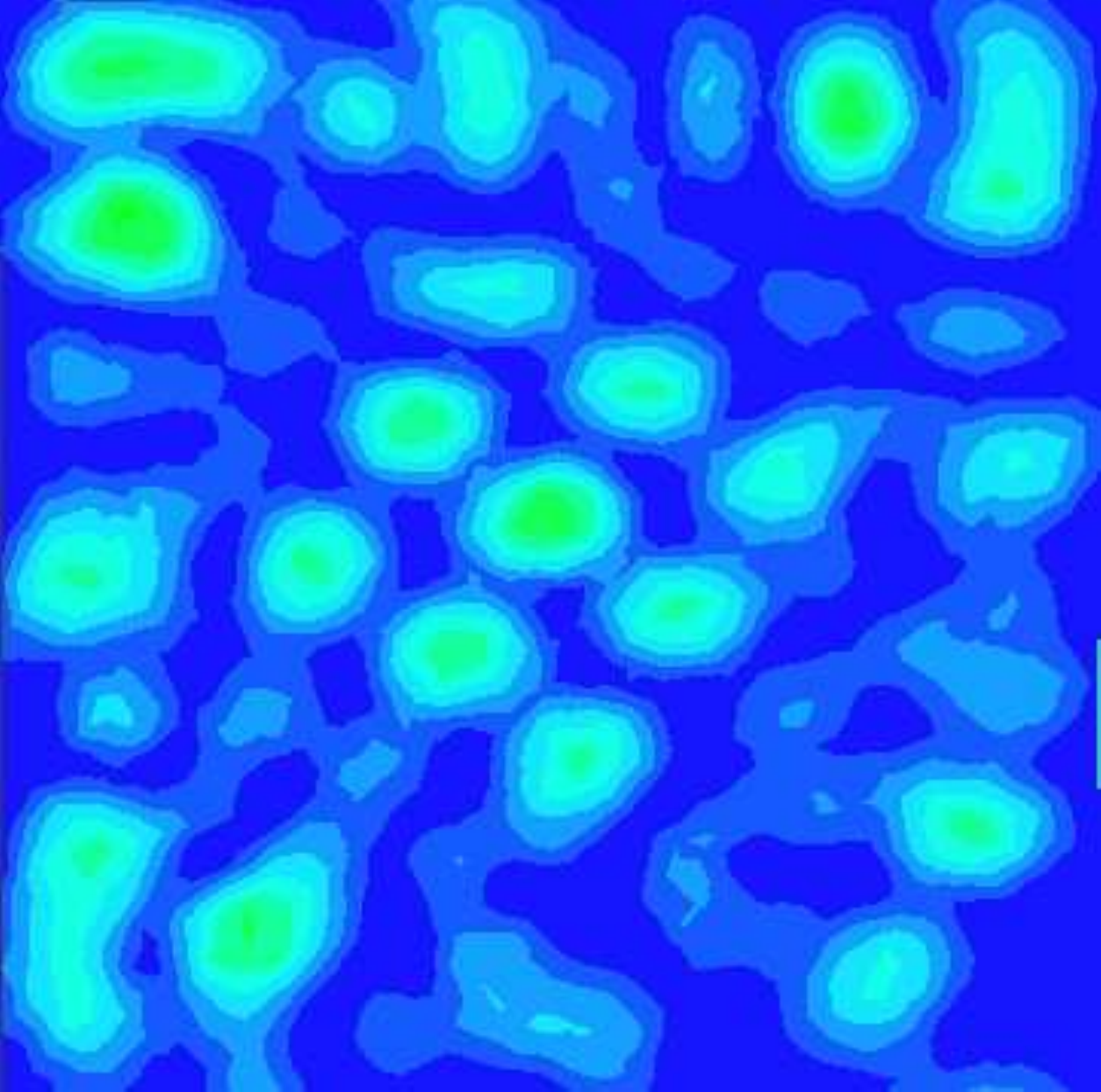}}
\centerline{$l=4$}
\end{minipage}

\centering
\begin{minipage}[t]{0.32\linewidth}
\centering
\centerline{\includegraphics[width=2.8cm,height=2.8cm]{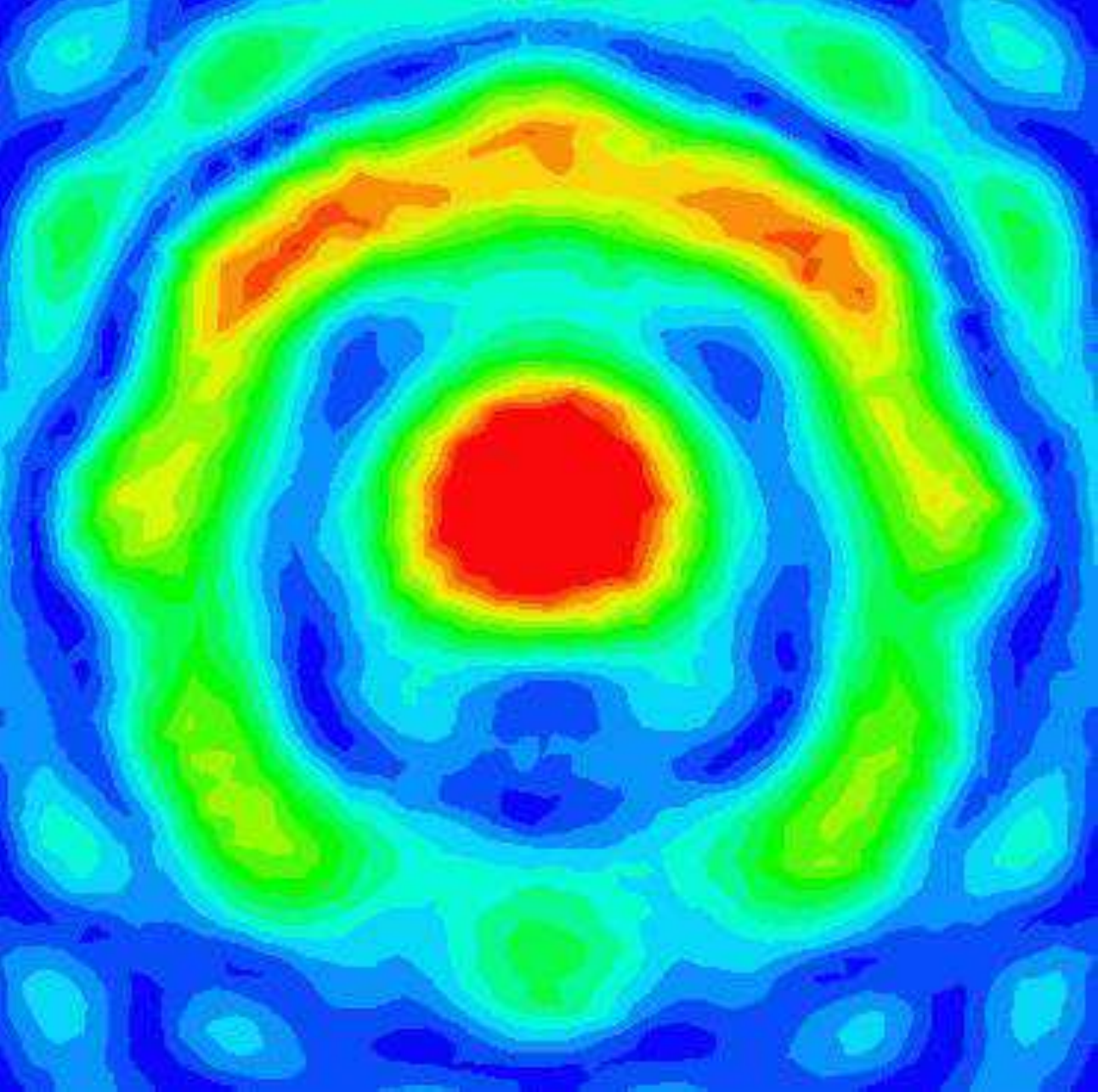}}
\centerline{$l=0$ (PE beam)}
\end{minipage}
\centering
\begin{minipage}[t]{0.32\linewidth}
\centering
\centerline{\includegraphics[width=2.8cm,height=2.8cm]{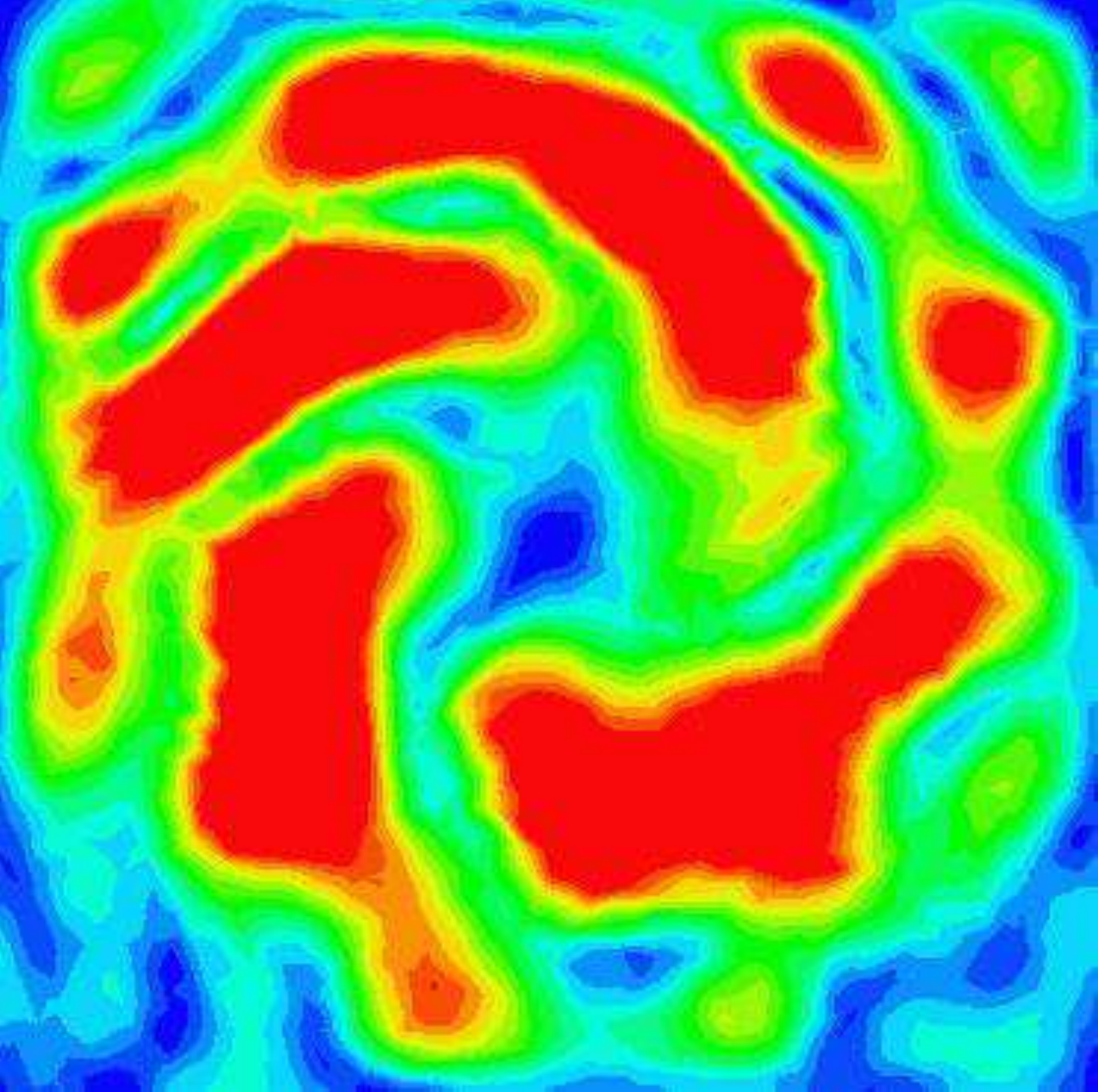}}
\centerline{$l=2$}
\end{minipage}
\centering
\begin{minipage}[t]{0.32\linewidth}
\centering
\centerline{\includegraphics[width=2.8cm,height=2.8cm]{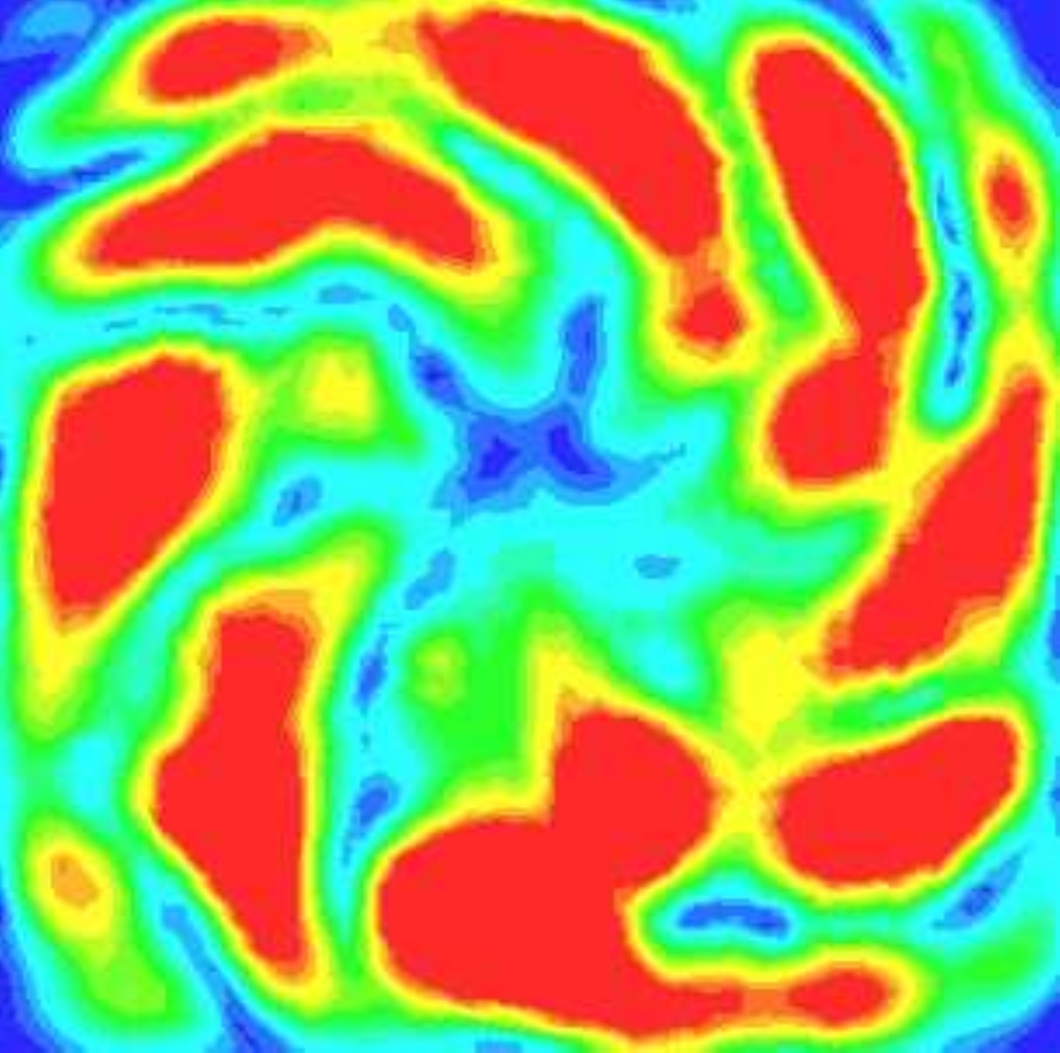}}
\centerline{$l=4$}
\end{minipage}
 \caption{E field of OAM beams (vertical direction).}\label{E_above_field}
  \vspace{-5pt}
\end{figure}

Figures~\ref{E_field} and~\ref{E_above_field} depict the E field of OAM beams observing from the horizontal direction and vertical direction, respectively. It can be seen from Fig.~\ref{E_field} that the divergent angle of OAM beam increases as the index of OAM-mode increases. When the propagation distance is very large, OAM beams of high-order OAM-mode diverge to be centrally hollow, which makes the radio vortex signal very difficult to be received. After converging by the bifocal lens antenna, the divergent degree of OAM beams is significantly reduced. Observing Fig.~\ref{E_field}, we can find that OAM beams are nearly converged into columnar beams, where we place the plane at $100\rm~mm$ to observe the results. The external lens does not cover the divergent angle of the side lobes. Thus, the side lobes of high-order OAM beams doesn't impact the transmission. For the plane beam, the gain of side lobes is downgraded due to very large power attenuation in the center of lens antenna. As shown in Fig.~\ref{E_above_field}, the E field of divergent OAM beams cannot be detected while the converged OAM beams can be easily detected on the height of $100\rm~mm$ above the UCA. We can also see that the convergent OAM beams still have the vortex wavefront characteristics, which confirms that lens antennas do not violate the wavefront characteristics of OAM beams.

\begin{figure*}[htbp]
 %\vspace{-5pt}
		\subfigure[Original]{
			\centering
			 \includegraphics[width=.3\textwidth,height=.225\textheight]{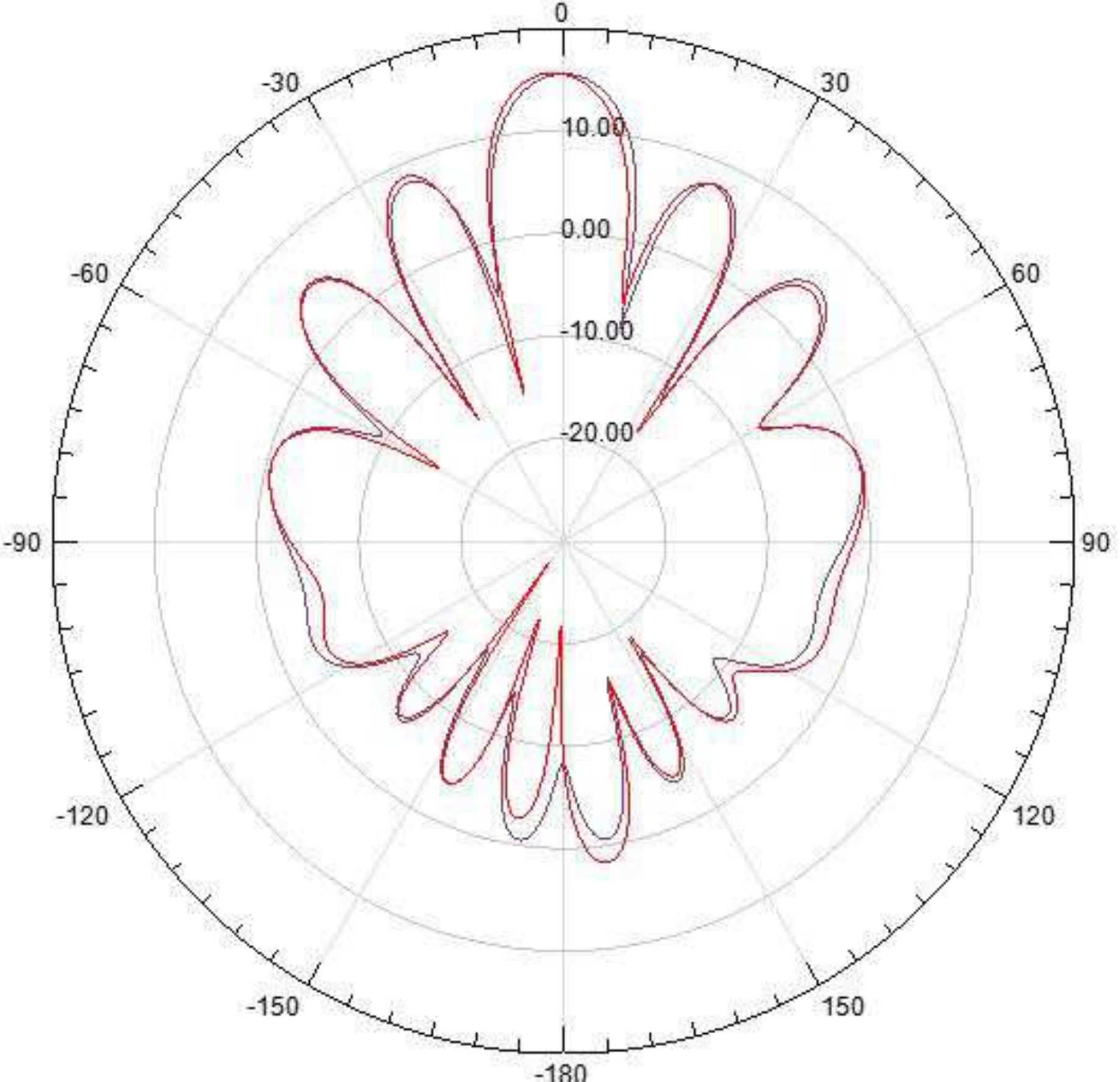}
			
		}
\subfigure[Lens converged]{
			\centering
			
			 \includegraphics[width=.3\textwidth,height=.225\textheight]{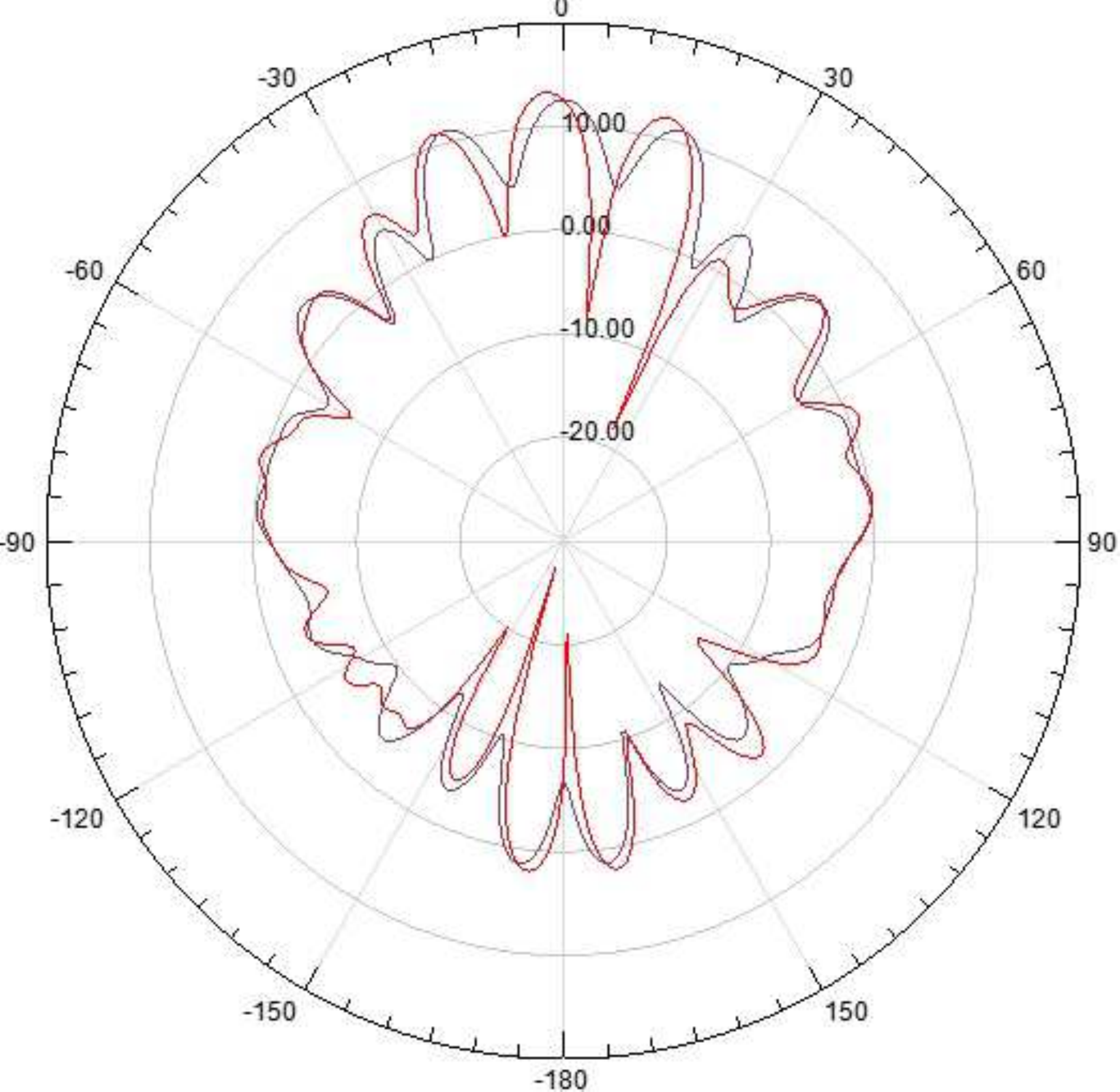}
			
		}
\subfigure[Bifocal lens converged]{
			\centering
			
			 \includegraphics[width=.3\textwidth,height=.225\textheight]{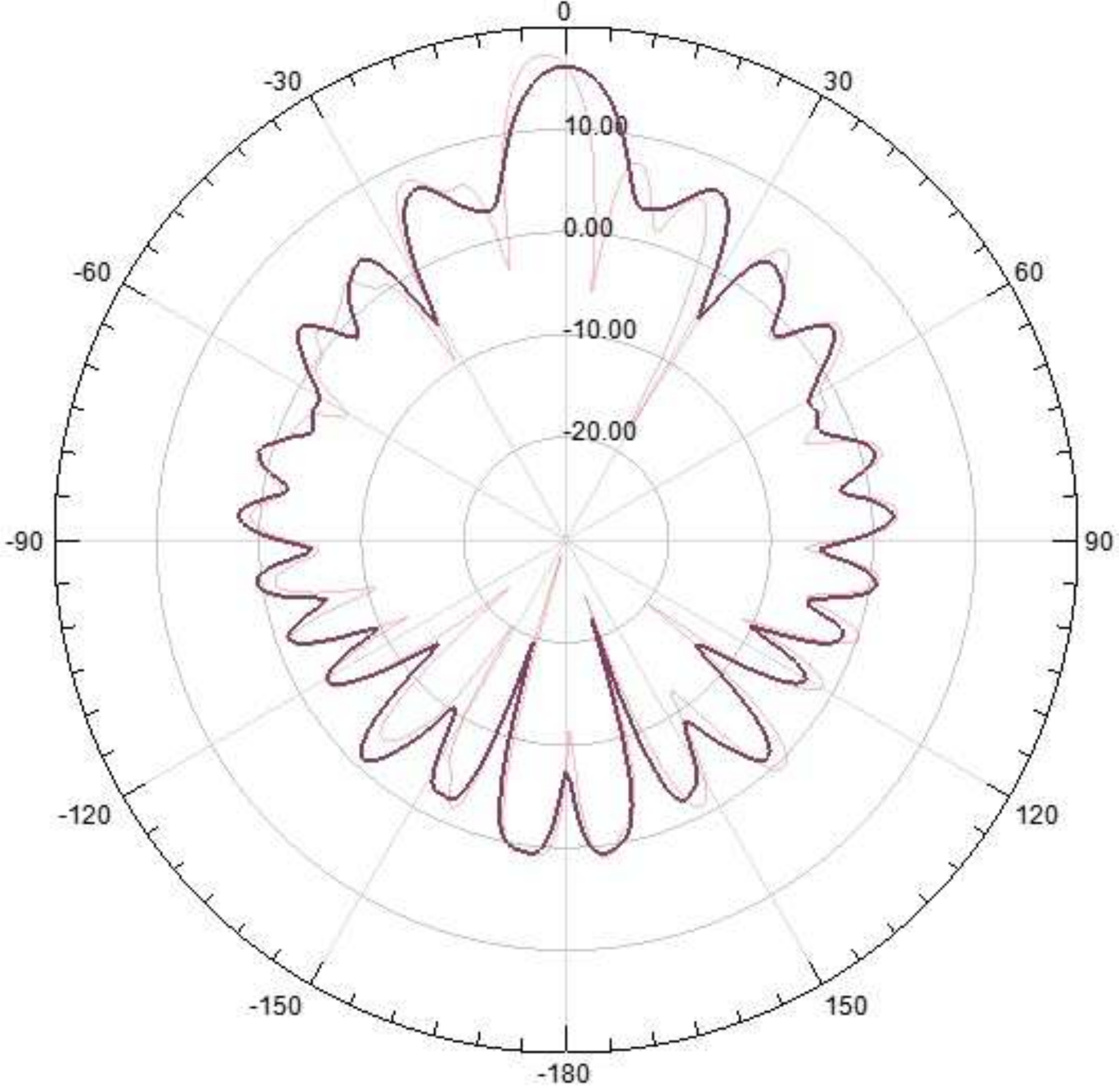}
		}

   \caption{2D radiation pattern of original PE beams, lens converged PE beams, and bifocal focal lens converged PE beams.}\label{2d_polar_comparison}
\end{figure*}

Figure~\ref{2d_polar_comparison} shows the 2D radiation pattern of original PE beams, lens converged PE beams, and bifocal focal lens converged PE beams. Adjacent OAM-modes have too much overlapped area with each other, which makes receivers very hard to separate them. Thus, we choose to simultaneously generate PE beam and vortex beams with OAM-mode 2 and 4, which satisfies the requirement of multiple independent date traffic transmissions. As shown in Fig.~\ref{2d_polar_comparison}, the gain of plane beam converged by lens antenna is severely reduced. To solve this problem, we reduce the thickness of the center of lens antenna. According to the simulation results, the downgrading is very small for the convergent plane beam when using the bifocal lens antenna.

% But the low-frequency beams have a much bigger beamwidth, and cannot be as concentrated as the light beams. And for the low mode, due to the strong convergence effect of the lens antenna, the side lobe is also convergence, resulting in a bigger energy than the main lobe. Moreover, the converged side lobe of lower modes is overlapped with the main lobe of high modes, which will disrupt the multiple modes transmission. Besides, mode 1 has too much overlapped area with plane wave, which makes it different for receivers to separate them. Observe the high mode converged beams, due to the high modes have a greater divergence angle, so there is a larger area to accommodate the wide beam. Therefore, it can be seen from the mode 2,3 that the convergence effect of OAM beams is great. Besides, the side lobe of high OAM modes would not be converged. Thus the side lobe will have little influence on the multiple modes transmission. However, the plane wave faces a huge decay due to the thickness of the lens antenna's center. And the side lobe of the plane wave interferes the high mode OAM beams.
\par %Comparing the simulation of bifocal lens antenna and single focal lens antenna, the convergence effect of the bifocal lens on the low mode is obviously weakened, and the convergence of the low mode side lobe is reduced. As can be seen from the plane beam, the side lobe is not overly converged, and the main lobe is more concentrated and has less decay. As shown in Fig.~\ref{E_field}, the plane wave's side lobe is interfere with a part of the main lobe, thus the side lobe is inhibited. 2 and 3 mode OAM beams are converged into columnar beams, thus simply the receiving at the receiving end. After adjustment of bifocal lens antenna, the convergence effect is much better than single focal lens antenna, as shown in Fig.~\ref{2d_polar_comparison}.
%\par According to our observation, the larger focal distance can reduce different modes' overlap caused by beam divergence. However, in order to cover larger modes, the diameter of the lens antenna should be increased accordingly, which makes the lens more bulky. Moreover, beam asymmetry may be caused by UCA's design and HFSS's simulation, which is not our concern in this paper.

%由仿真图可以明显看出，未汇聚过的涡旋波随着模态增加发散角度逐渐增大。随着传输距离的增加，高模态的涡旋波会大幅度发散，接收天线很难接收到。

%单透镜汇聚后，涡旋波的发散程度明显减小。但是由仿真图明显可以看到，虽然每个模态有固定最大的角度。

%但是从截面图可以看出，涡旋波是有一定自身衍射的，
%无法像光波的能量一样集中，所以由于衍射，部分模态的涡旋波在汇聚后会产生混叠。而且对于低模态，由于透镜的汇聚作用较强，传输中不需要的旁瓣也被汇聚，
%会严重影响正常模态的涡旋波的传输，甚至能量高于原本的低模态。如图所示，0模态和1模态由于汇聚效果过强旁瓣的能量被集中，反而超越主模态，这会严重影响接收。
%在多模态同时发送时，这总混叠现象会使本来相互正交的不同模态波混在一起，难以实现多模态传输。\\
%观察单透镜汇聚的高模态，由于高模态发射角度较大，所以有更大角度空间
%容纳低频波束的衍射，因此可以从2、3、4模态中看出，汇聚效果很明显，有$5^{\circ}-10^{\circ}$的汇聚。而且在高模态，旁瓣没有被汇聚，所以会使发射的波束减少旁瓣的影响，

%主瓣明显减弱。0 1模态的汇聚减弱，防止两者混叠在一起。双焦距透镜对高模态的汇聚进一步加强，对2、3 模态汇聚超过$10^{\circ}$，对4模态的汇聚超过$15^{\circ}$，说明双焦透镜对不同模态
%有着不同的汇聚作用。对低模态的汇聚作用较小，防止因为低频波束的衍射产生混叠；对高模态的汇聚加强，使高模态发散进一步降低。内外透镜相互配合，能够保证不同模态的波束在互不干扰的
%前提下尽可能汇聚，从而实现多模态传输。

%焦距大可以减轻由波束发散导致的不同模态重叠现象。但是为了实现对大模态汇聚，透镜的尺寸也要相应增加。

%(波束不对称可能是由UCA导致的)

\begin{figure}
\centering
%\vspace{-10pt}
\includegraphics[scale=0.33]{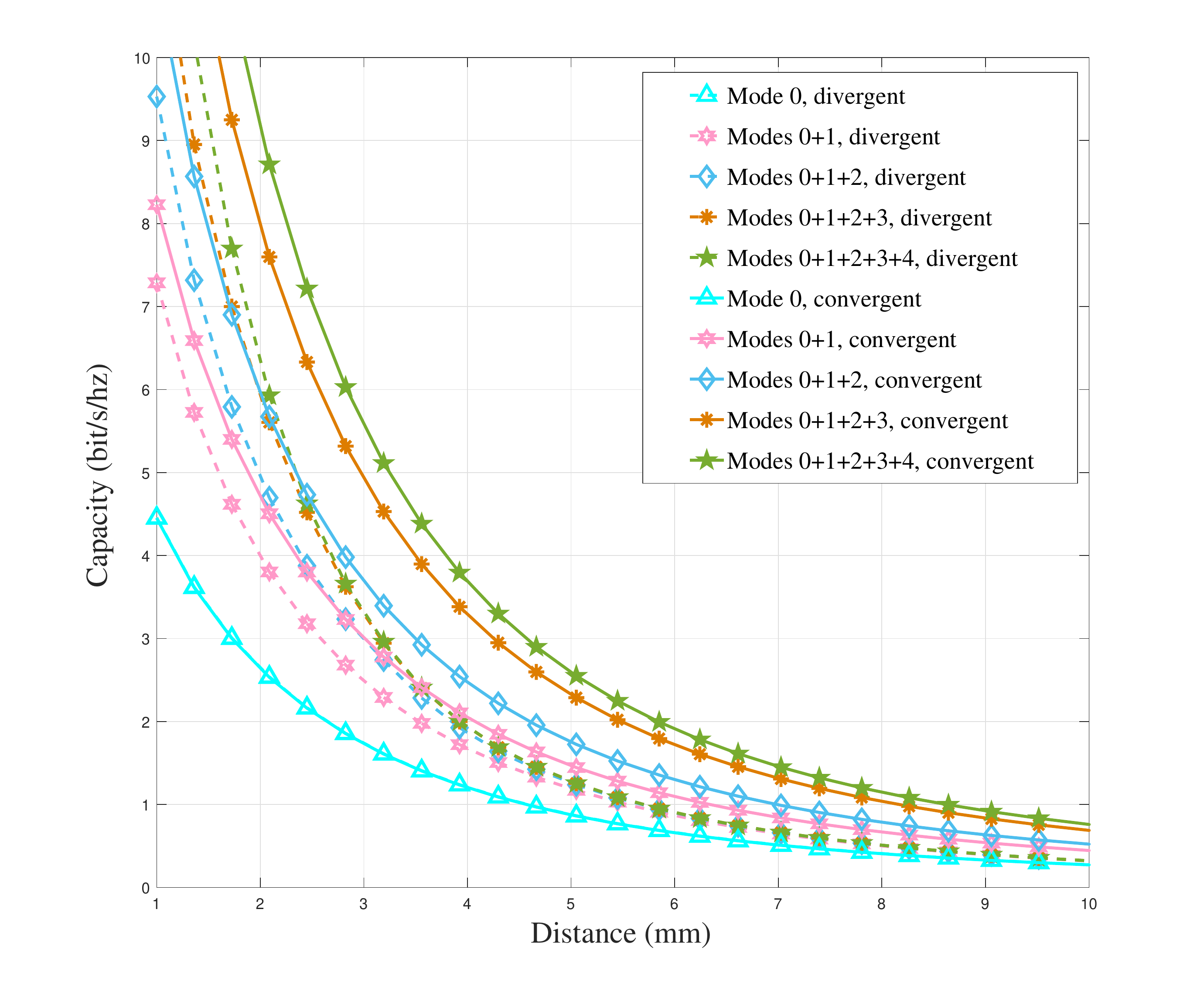}
\caption{Capacities of OAM beams based communication.}\label{fig:capacity}
\vspace{-10pt}
\end{figure}

Figure~\ref{fig:capacity} shows the capacities of OAM beams based wireless communication. As shown in Fig.~\ref{fig:capacity}, the capacity of convergent OAM beam based radio vortex wireless communication using bifocal lens is larger than that of the divergent OAM beam based radio vortex wireless communication. For the radio vortex communication using OAM-mode 1, the capacity of divergent OAM based radio vortex wireless communication is very close to the capacity of converged OAM based radio vortex wireless communication. This result shows that the bifocal lens does not downgrade the beams too much while the convergence makes the power more concentrated. Moreover, the advantage of convergent OAM based radio vortex communication significantly increases as the number of OAM-mode increases. For the divergent OAM based radio vortex communication, the incremental of capacity reduces as the OAM-mode increase because the high-order OAM-mode can be received with low power. Although the lens causes somewhat SNR downgrading, the convergence makes the power more concentrated. Thus, the capacity of convergent OAM beam based radio vortex wireless communication is significantly increased.
\par
\begin{figure}
\centering
%\vspace{-10pt}
\includegraphics[scale=0.33]{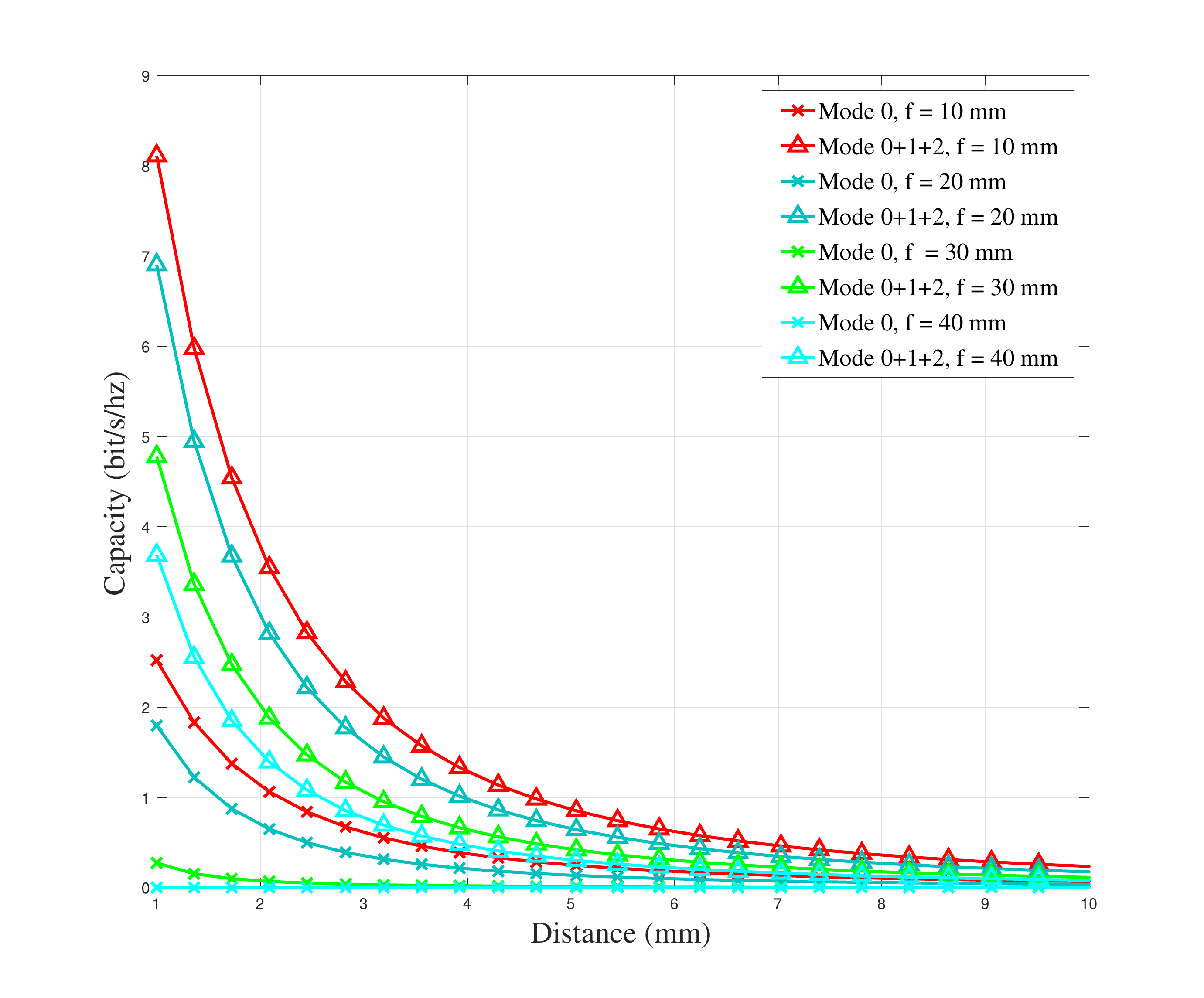}
\caption{Capacities of OAM beams based communication versus focal distance with focal lens.}\label{fig:CapacityWithFocal}
\vspace{-10pt}
\end{figure}

 Figure~\ref{fig:CapacityWithFocal} shows the capacities of OAM beams based wireless communication taking into account the focal distance of lens antenna. The focal distance and diameter impact the thickness of lens. When the diameter of lens remains unchanged, the lens becomes thinner as the focal distance increases. However, in our lens design, in order to converge multiple OAM beams, the lens should be able to cover the maximum divergent angle of OAM beams. The diameter of lens is in associate with the focal distance. The thickness of lens grows as focal distance increases, thus causing a severe downgrading for OAM beams. Therefore, the capacity decreases as focal distance increases. In addition, the divergent angles of OAM beams with different OAM-modes are different. Table.~\ref{tab:R_theta} shows that the divergent angle increases as OAM-mode increases. The center of lens is thicker, causing greater downgrading. It can be also seen from Fig.~\ref{fig:CapacityWithFocal} that when $f=40 \rm~ mm$, the capacity of OAM-mode 0 is very close to zero. This is because the received SNR is very close to zero given the fixed reception area.
\par
\begin{figure}
\centering
%\vspace{-10pt}
\includegraphics[scale=0.33]{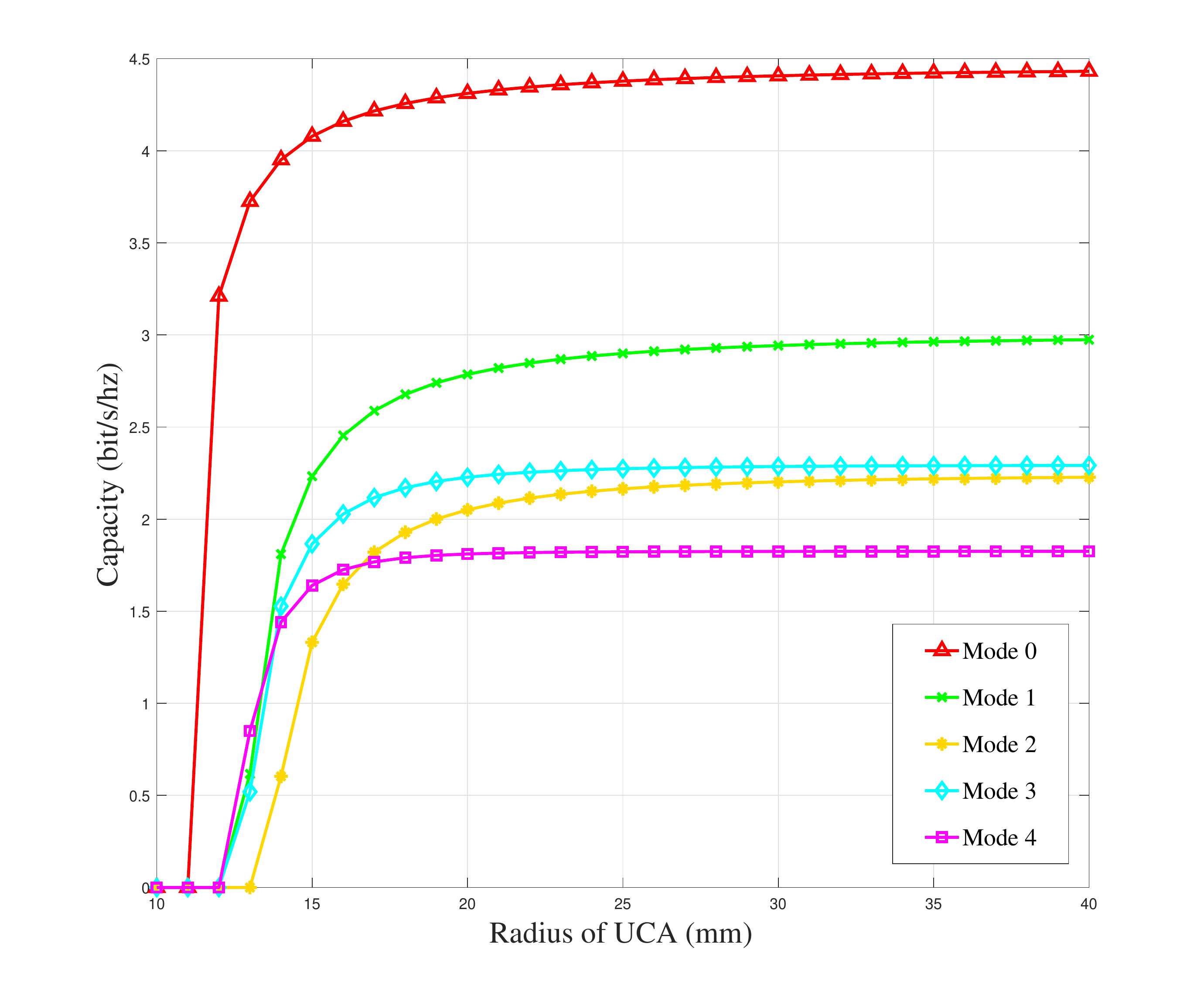}
\caption{Capacities of OAM beams based communication with different radius of UCA.}\label{fig:CapacityWithR}
\vspace{-10pt}
\end{figure}

 Figure~\ref{fig:CapacityWithR} shows the capacities of OAM beams based wireless communication with different radius of UCA. The maximum divergent angle increases with the radius decreases. The diameter of lens increase as the increase of radius. Thus, The center of lens is thinner, reducing the SNR downgrading. We can conclude from Fig~\ref{fig:CapacityWithR} that the capacity of OAM beams based wireless communication increases as the increase of radius of UCA. However, as the radius becomes bigger, the capacity grows at a decreasing rate. Thus, for the sake of the strength of OAM beams, the radius should be selected in the middle range. In addition, according to our simulation, $a$ in Eq.~\eqref{eq:E_after_lens} is about $10^{-3}$. Thus, the SNR downgrading caused by lens has a much greater effect than the energy redistribution of lens.

\section{Conclusions}\label{sec:conc}
In this paper, we proposed the UCA and lens based antennas to converge OAM beams of multiple OAM-modes. We obtained that multiple OAM beams generated by UCA antenna can be converged by lens antennas, while the wavefront property of OAM beams can be maintained. Moreover, our proposed bifocal lens can significantly decrease the downgrading on the strength of PE beam and OAM beams with low-order OAM-mode. We proved that after converging, the capacity of radio vortex wireless communications can be significantly increased. In addition, we give the relationship between the divergent angle of OAM beams and the radius of UCA antenna, which provides the guidance for the design of UCA antenna.

\appendix
 \section{Appendix}
 \subsubsection{Design of UCA Antenna}
\par We design the patch based UCA antenna. The profile of a 16 patch elements based UCA is shown in Fig.~\ref{fig:UCA}, where patch elements are equally distributed around the circle. To achieve the best feed performance, it is needed to design the size of patch elements~\cite{Milligan1985Modern}. The width of the patch, denoted by $W_{P}$, is given as follows~\cite{Bahl1980Microstrip,Bahl1982Design}:
%In order to obtain the vortex beam with high index of OAM-mode, 16

\begin{eqnarray}
W_{P}=\frac{c}{2f_{r}}\left(\frac{\varepsilon_r+1}{2}\right)^{-1/2},
\label{eq:patch_W}
\end{eqnarray}
where $c$ represents the speed of light in vacuum, $\varepsilon_{r}$ is relative permittivity, and $f_{r}$ is the operating frequency of antenna.

Then, we can obtain the relative effective permittivity of the dielectric substrate, denoted by $\varepsilon_{re}$, as follows: %确定介质基板的相对有效介电常数
\begin{eqnarray}
\varepsilon_{re}=\frac{\varepsilon_{re}+1}{2}+\frac{\varepsilon_{re}-1}{2}\bigg(1+\frac{10h}{W_P}\bigg)^{-1/2},
\label{eq:patch_re}
\end{eqnarray}
where $h$ is the thickness of dielectric substrate.

The patch element edge field results in the equivalent length of radiation gap (stretching effects)~\cite{Bahl1980Microstrip}, denoted by $\Delta L$, which is expressed as follows:
\begin{eqnarray}
\Delta L=0.412h\frac{(\varepsilon_{re}+0.3)(W_P/h+0.264)}{(\varepsilon_{re}-0.258)(W_P/h+0.8)}.
\label{eq:Delta L}
\end{eqnarray}
Theoretically, the length of patch element, denoted by $L_P$, should be half of the effective wavelength. However, due to the stretching effects of the field,
the stretching length $\Delta L$ needs to be removed from the patch element's length $L_P$. Thus, we have:
%矩形贴片天线的长度$L_P$理论上应取有效波长的一半，由于边缘场影响，
%故拉伸长度$\Delta L$应从长度中去掉，定义有效波长为$\lambda_g$：
\begin{eqnarray}
L_P=\frac{\lambda_g}{2}-2\Delta L=\frac{c}{2f_{r}\sqrt{\varepsilon_{re}}}-2\Delta L,
\label{eq:patch_l}
\end{eqnarray}
where $\lambda_g$ is the effective wavelength~\cite{Javor1995Design}.

\bibliographystyle{IEEEbib}
\bibliography{References}

\end{document}